\documentclass[12pt]{article}
\usepackage{geometry}
\usepackage{amsmath}
\usepackage{amssymb}
\usepackage{amsmath}
\usepackage{amsthm}
\usepackage{bbold}
\usepackage{xspace}
\usepackage{graphicx}

\def\appendix#1{
  \addtocounter{section}{1}
  \setcounter{equation}{0}
  \renewcommand{\thesection}{\Alph{section}}
 \section*{Appendix \thesection\protect\indent \parbox[t]{11.715cm} {#1}}
  \addcontentsline{toc}{section}{Appendix \thesection\ \ \ #1}
  }

\renewcommand{\thefootnote}{\fnsymbol{footnote}}

\numberwithin{equation}{section}

\def\pint{{-\!\!\!\!\!\!\int}\,}

\newcommand{\be}{\begin{equation}}
\newcommand{\ee}{\end{equation}}
\newcommand{\ba}{\begin{aligned}}
\newcommand{\ea}{\end{aligned}}

\newcommand{\ie}{{\it i.e.}}

\newcommand{\sltwo}{\mathfrak{sl}(2)}
\newcommand{\sutwo}{\mathfrak{su}(2)}
\newcommand{\psu}{\mathfrak{psu}}

\makeatletter
\def\sla@#1#2#3#4#5{{%
  \setbox\z@\hbox{$\m@th#4#5$}%
  \setbox\tw@\hbox{$\m@th#4#1$}%
  \dimen4\wd\ifdim\wd\z@<\wd\tw@\tw@\else\z@\fi
  \dimen@\ht\tw@
  \advance\dimen@-\dp\tw@
  \advance\dimen@-\ht\z@
  \advance\dimen@\dp\z@
  \divide\dimen@\tw@
  \advance\dimen@-#3\ht\tw@
  \advance\dimen@-#3\dp\tw@
  \dimen@ii#2\wd\z@  \raise-\dimen@\hbox to\dimen4{%
    \hss\kern\dimen@ii\box\tw@\kern-\dimen@ii\hss}%
  \llap{\hbox to\dimen4{\hss\box\z@\hss}}}}
\def\slashed#1{%
  \expandafter\ifx\csname sla@\string#1\endcsname\relax
    {\mathpalette{\sla@/00}{#1}}%
  \else
    \csname sla@\string#1\endcsname
  \fi}
\makeatother

\begin{document}


\thispagestyle{empty}
\begin{flushright}\footnotesize
\texttt{hep-th/0507189}\\
\texttt{AEI-2005-125}\\
\texttt{DESY-05-108}\\
\texttt{ITEP-TH-49/05}\\
\texttt{UUITP-11/05}\\
\texttt{ZMP-HH/05-12}\\
\vspace{0.8cm}
\end{flushright}

\renewcommand{\thefootnote}{\fnsymbol{footnote}}
\setcounter{footnote}{0}

\begin{center}
{\Large\textbf{\mathversion{bold}
Quantum corrections to spinning strings in $AdS_5 \times S^5$ and
Bethe ansatz: a comparative study \\
}\par}

\vspace{1.5cm}

\textrm{Sakura Sch\"afer-Nameki$^{\alpha}$, Marija
Zamaklar$^{\beta}$ and Konstantin Zarembo$^\gamma$\footnote{Also at
ITEP, Moscow, Russia}} \vspace{8mm}

\textit{$^{\alpha}$ II. Institut f\"ur Theoretische Physik der Universit\"at Hamburg\\
Luruper Chaussee 149, 22761 Hamburg, Germany} \\
\texttt{sakura.schafer-nameki@desy.de}
\vspace{3mm}

\textit{$^{\alpha}$ Zentrum f\"ur Mathematische Physik, Universit\"at Hamburg\\
Bundesstrasse 55, 20146 Hamburg, Germany} \vspace{3mm}

\textit{$^{\beta}$ Max-Planck-Institut f\"ur Gravitationsphysik, AEI\\
Am M\"uhlenberg 1, 14476 Golm, Germany}\\
\texttt{marzam@aei.mpg.de}
\vspace{3mm}

\textit{$^{\gamma}$ Department of Theoretical Physics, Uppsala University\\
751 08 Uppsala, Sweden}\\
\texttt{Konstantin.Zarembo@teorfys.uu.se}
\vspace{3mm}


\par\vspace{1cm}

\textbf{Abstract}\vspace{5mm}

\end{center}
We analyze quantum corrections to rigid spinning strings in
$AdS_5\times S^5$. The one-loop worldsheet quantum correction to the
string energy is compared to the finite-size correction from the
quantum string Bethe ansatz. Expanding the summands of the string
theory energy shift in the parameter $1/{\mathcal J}^2$ and
subsequently resumming them yields a divergent result. However, upon
zeta-function regularisation these results agree with the Bethe
ansatz in the first three orders. We also perform an analogous
computation in the limit of large winding number, which results in a
disagreement with the string Bethe ansatz prediction. A similar
mismatch is observed numerically. We comment on the possible origin
of this discrepancy.


\vspace*{\fill}

\newpage
\setcounter{page}{1}
\renewcommand{\thefootnote}{\arabic{footnote}}
\setcounter{footnote}{0}


\tableofcontents

\newpage

\section{Introduction}

Understanding the quantum spectrum of string theory in $AdS_5 \times
S^5$ is an important open problem. Solving this problem will open up
venues for testing the ideas of gauge/string duality in the genuine
stringy regime. It is becoming more and more clear that progress in
quantizing strings on $AdS_5 \times S^5$ is impossible without
serious input from the dual ${\mathcal N}=4$ supersymmetric
Yang-Mills theory (SYM). One idea that has proved extremely useful
on the gauge theory side and could potentially be applied to $AdS$
strings, is to compute the spectrum using a Bethe ansatz. The Bethe
ansatz is the standard approach to quantize integrable systems
\cite{Faddeev:1996iy} and it is believed that both planar ${\mathcal
N}=4$ SYM and string theory in $AdS_5\times S^5$ are integrable.

As was observed first at one loop \cite{Minahan:2002ve,
Beisert:2003yb} and then at higher orders in perturbation theory
\cite{Beisert:2003tq,Beisert:2003ys,Serban:2004jf}, the planar
dilatation operator of ${\mathcal N}=4$ SYM can be identified with a
Hamiltonian of an integrable spin chain\footnote{Although the
dilatation operator is not integrable beyond leading order in
the $1/N$ expansion \cite{Beisert:2003tq}, the planar integrability
is still useful in the study of decays of semiclassical strings
\cite{Peeters:2004pt} and in the computation of three-point
functions \cite{Okuyama:2004bd}. We should also mention that the
classical equations of motion of ${\mathcal N}=4$ SYM admit a Lax
representation \cite{Volovich:1984ra}, but we do not know if this
property has anything to do with the quantum integrability of the
planar dilatation operator.}. The integrability on the string theory
side arises because the classical world-sheet sigma-model admits a
Lax representation. For the bosonic reduction this almost
immediately follows \cite{Mandal:2002fs} from the integrability of
the $O(n)$ model \cite{Pohlmeyer:1975nb}.  The Lax pair for the full
supersymmetric sigma-model in $AdS_5\times S^5$
\cite{Metsaev:1998it} was constructed in \cite{Bena:2003wd}.

Because the classical equations of motion of the $AdS$ string are
integrable, their solutions can be parameterized by the spectral
data of the Lax operator. By reformulating the standard solution of
the spectral problem \cite{books} it was shown in
\cite{Kazakov:2004qf} that the spectral density for the string
moving on the $\mathbb{R}\times S^3$ subspace of $AdS_5\times S^5$
satisfies an integral equation that strikingly resembles the
large-volume (thermodynamic) limit of the quantum Bethe equations
for the spectrum of the dilatation operator in the dual gauge
theory. These results were extended to other sectors
\cite{Kazakov:2004nh,Beisert:2004ag,Schafer-Nameki:2004ik,Alday:2005gi}
and eventually to the most general solution including world-sheet
fermions \cite{Beisert:2005bm}. Of course the classical
approximation in the sigma-model is accurate only at strong 't~Hooft
coupling (\ie\ weak worldsheet coupling). In addition, the Noether
charges of the string have to be large. In order to quantize the
string one needs to ``undo'' the thermodynamic limit and turn the
integral equations for the sigma-model into discrete, quantum string
Bethe equations. Such a discretization was first proposed for the
$\sutwo$ subsector \cite{Arutyunov:2004vx}, then for other rank-one
sectors \cite{Staudacher:2004tk} and subsequently for the complete
set of Bethe equations with the $\psu(2,2|4)$ symmetry
\cite{Beisert:2005fw}. The quantum string Bethe equations work
remarkably well in several tractable limits:
they have the right classical limit (by construction), reproduce the
leading quantum corrections for the BMN states and yield the correct
energies of massive states in the strict strong-coupling limit.

There are very few explicit calculations for quantum strings in
$AdS_5\times S^5$. One major example is string quantization in the
plane-wave limit \cite{Berenstein:2002jq} which leads to a solvable
string theory \cite{Metsaev:2001bj} and  can be understood as
quantization around the simplest point-like solution of the string
spinning on $S^5$ \cite{Gubser:2002tv}. The curvature corrections
\cite{Parnachev:2002kk} to the string states in this background (BMN
states) were calculated in \cite{Callan:2004uv}. Frolov-Tseytlin
solutions \cite{Frolov:2003qc,Tseytlin:2003ii} generalize this setup
to macroscopic strings and it is possible to quantize fluctuations
around these solutions in some cases
\cite{Frolov:2003tu,Frolov:2004bh,Park:2005ji,Fuji:2005ry}. For
these solutions, the classical string energies can be compared to
the anomalous dimensions in the gauge theory (see
\cite{Tseytlin:2003ii,Beisert:2004ry} for review), because
the 't~Hooft coupling $\lambda $  combines with the R-charge $J$
into the BMN coupling $1/\mathcal{J}^2\equiv \lambda /J^2$, which
can be small even if the 't~Hooft coupling is large, provided that
the R-charge is large enough. In particular,
the string action reduces to the effectve action of the spin chain
in the limit of large $\mathcal{J}$ \cite{Kruczenski:2003gt}.
Generically, one finds that string
theory and SYM agree up to two loops and start to disagree at three
loops. For the quantum corrections the comparison has only been done
at the one-loop level \cite{Beisert:2005mq, Hernandez:2005nf}.
It would be interesting to understand what happens at higher orders
of perturbation theory.



Our goal is to compare quantum corrections to macroscopic strings
with the quantum string Bethe ansatz at higher loops
\cite{Arutyunov:2004vx,Staudacher:2004tk,Beisert:2005fw}. The
conjectured quantum string Bethe equations were rigorously tested at
infinite $\lambda $, but they can potentially receive
$1/\sqrt{\lambda }$ corrections \cite{Arutyunov:2004vx}. Comparison
of the quantum string Bethe ansatz
to the direct quantum string calculation provides an explicit
check of whether such corrections are present at
$O(1/\sqrt{\lambda })$ or not. Furthermore, the string Bethe
equations are known to exactly reproduce the first two orders of the
SYM perturbation theory independently of $J$ \cite{Beisert:2004jw},
and we can just expand the energies computed from them in the
't~Hooft coupling to find the two loop anomalous dimensions in SYM.
In this way we can extend the analysis of
 \cite{Beisert:2005mq, Hernandez:2005nf} to two loops.

Let us briefly review the classical string configurations that we
shall study. The one-loop quantum corrections were computed for two
classes of string solutions -- for circular strings rotating in
$S^5$ with two independent angular momenta
\cite{Frolov:2003tu,Frolov:2004bh} and for circular strings spinning
in $AdS_3$ and rotating around $S^5$ \cite{Park:2005ji}. The first
case is plagued by instabilities \cite{Frolov:2003qc,Frolov:2003tu}
and for this reason we shall concentrate on strings moving in
$AdS_3\times S^1\subset AdS_5\times S^5$ \cite{Arutyunov:2003za}
(throughout the paper, we shall adopt the conventions of
\cite{Park:2005ji}). The relevant part of the $AdS_5\times S^5$
metric in global coordinates is
\begin{equation}\label{}
 ds^2=-\cosh^2\rho \,dt^2+d\rho ^2+\sinh^2\rho \,d\theta  ^2
 +d\phi  ^2,
\end{equation}
where the first three terms are the metric of $AdS_3$ and $\phi $ is
the angle of a big circle in $S^5$. The circular string solution has
the following form
\begin{equation}\label{classt}
 \rho =\,{\rm const}\,,\qquad t=\kappa \tau ,\qquad
 \theta =\sqrt{\kappa ^2+k^2}\,\tau +k\sigma ,\qquad
 \phi =\sqrt{\kappa ^2+m^2}\,\tau +m\sigma ,
\end{equation}
where
\begin{eqnarray}\label{ura}
 r_1^2\equiv\sinh^2\rho &=&\frac{\mathcal{S}}{\sqrt{\kappa ^2+k^2}}\,,
 \\
\label{ura1}
\mathcal{E}&=&
 \frac{\kappa \mathcal{S}}{\sqrt{\kappa ^2+k^2}}+\kappa,
 \\ \label{ura2}
 2\kappa \mathcal{E}-\kappa ^2&=&2\sqrt{\kappa ^2+k^2}\,\mathcal{S}
 +\mathcal{J}^2+m^2\,, \\
 k\mathcal{S}+m\mathcal{J}&=&0.
\end{eqnarray}
Global charges of the string (the energy $E$, the spin $S$, and the
angular momentum $J$) combine with the  string tension into the
following ``dimensionless" ratios, which stay finite in the classical
($\lambda \rightarrow \infty$, $J \rightarrow \infty$, $S\rightarrow
\infty$) limit \cite{Tseytlin:2003ii}:
\begin{equation}\label{char}
 \mathcal{E}=\frac{E}{\sqrt{\lambda }},
 \qquad \mathcal{S}=\frac{S}{\sqrt{\lambda }},
 \qquad \mathcal{J}=\frac{J}{\sqrt{\lambda }} \, .
\end{equation}
Thus $1/\sqrt{\lambda }$ or $1/J$ can be used interchangeably as the
loop counting parameters in the sigma-model. In addition, at any given
order in $1/J$ one can further expand in the BMN coupling
$1/\mathcal{J}^2=\lambda /J^2$. In this way one recovers the two-loop
perturbative SYM results.

In section 2 we review the string theory computation and evaluate
the energy shift, at leading order in $1/J$ and at the first three orders
in $1/\mathcal{J}^2$. Although the
exact energy shift is finite, individual terms of the
$1/\mathcal{J}$ expansion diverge. To render the results
finite  we use a particular
prescription, the zeta-function regularization.

In section 3 we compute the energy shift from the quantum string Bethe
ansatz, again perturbatively in $1/\mathcal{J}$. Unlike in the string
theory calculation, the $1/\mathcal{J}$ expansion is manifestly finite. However,
 the resulting expressions agree with the zeta-regularized string energy shift at
third order in perturbation theory.

In section 4 we calculate the energy shift in the non-perturbative
regime (\ie, small $\mathcal{J}$) of large winding number.
The energy shift is finite on both
sides in this case. We find a clear discrepancy between the Bethe
ansatz and the string calculation. In section 5, we present
numerical results which support the analytical evidence for the
discrepancy.

We discuss our results in section 6. Various technical details
 are collected in the appendices.


\section{Quantum corrections in string theory}

\subsection{Energy shift}

The semiclassical string quantization of \cite{Park:2005ji} yields
the following correction to the classical energy (\ref{ura1})
\begin{equation}
\label{full} \delta E^{string} = \delta E^{(0)}+ \delta
E^{osc} \, .
\end{equation}
Here the zero-mode contribution is given by
\begin{equation}\label{ZeroModeE}
\delta E^{(0)} = {1\over 2 \kappa} \left(4 \nu + 2 \kappa + 2
\sqrt{\kappa^2 + (1 + r_1^2) k^2} - 8  \sqrt{c^2 + a^2}\right).
\end{equation}
 The oscillator part has the following form
\begin{equation}\label{StringEnergyShift}
\begin{aligned}
\delta E^{osc} =&  {1\over \kappa} \sum_{n=1}^\infty
                   \bigg( 4 \sqrt{n^2 + \nu^2} + 2 \sqrt{n^2 + \kappa^2}
                   - 4\sqrt{(n+\gamma)^2 + \alpha^2} -4
                   \sqrt{(n-\gamma)^2 + \alpha^2}
                 \cr
                &  \quad \qquad
             +\frac{1}{2} \sum_{I=1}^4 sign (C_{I}^{(n)}) \omega_{I, n} \bigg)\,,
\end{aligned}
\end{equation}
where the last term is the contribution of the $\sltwo$-modes, which
are the four solutions of the quartic equation
\begin{equation}\label{PolyPTT}
 (\omega^2-n^2)^2+4r_1^2\kappa ^2\omega^2-4(1+r_1^2)
 \left(\sqrt{\kappa ^2+k^2}\,\omega-kn\right)^2=0.
\end{equation}
The first line corresponds to the transverse and fermionic modes.
The various parameters are defined as \be \ba \nu    &=
\sqrt{\mathcal{J}^2 - m^2} \cr \alpha &= \sqrt{\frac{\kappa^2 +
\nu^2}{2}} \cr r_1^2    &= {\kappa^2 - 2 m^2 - \nu^2 \over 2 k^2}= -
{m \over k} {\mathcal J \over \sqrt{\kappa^2 + k^2}}\cr \gamma &=
{1\over 2} \kappa \left(1+ {2 k^2 (1 + r_1^2) \over \kappa^2
-\nu^2}\right)
          \sqrt{{\kappa^2 - \nu^2 - 2 k^2 r_1^2\over 2 (\kappa^2 +
          k^2)}} \,.
\ea \ee
 The sign factors are determined from
 \begin{equation}\label{signs}
C_{I}^{(n)} = (\omega_{I,n}^2- n^2) \prod_{J\neq I} (\omega_I -
\omega_J) .
\end{equation}

It is possible to perform a partial summation of the series
(\ref{StringEnergyShift}). The series is absolutely convergent,
because the summand decreases as $1/n^2$ at $n\rightarrow \infty $.
Therefore one can sum each frequency separately by regularizing the
divergences; one adds and subtracts terms of the form $c_1n+c_2/n$
before separating various frequencies. This does not change the
result, because each partial sum is again absolutely convergent.
 The basic sum is
\begin{equation}\label{}
 \sum_{n=1}^{\infty }\left[\sqrt{(n+\gamma )^2+\alpha ^2}
 +\sqrt{(n-\gamma )^2+\alpha ^2}-2n-\frac{\alpha ^2}{n}\right]
 =\gamma ^2-\sqrt{\gamma ^2+\alpha ^2}+F(\{\gamma \},\alpha ),
\end{equation}
where $\{\gamma \}$ denotes the fractional part of $\gamma $ and the
function $F(\beta ,\alpha )$ is defined by the following integral
representation
\begin{equation}\label{defFbetaalpha}
 F(\beta ,\alpha )\equiv
 \sqrt{\alpha ^2+\beta ^2}-\beta ^2+\alpha ^2\int_{0}^{\infty }
 \frac{d\xi }{\,{\rm e}\,^\xi -1}\left(
 \frac{2J_1(\alpha \xi )}{\alpha \xi }\,\cosh \beta \xi -1
 \right).
\end{equation}
 Using
this result we find
\begin{eqnarray}\label{dEosc}
 \delta E^{osc} &=&\frac{1}{\kappa }\left[
 2F(0,\nu )+F(0,\kappa )-4F(\{\gamma \},\alpha )-2\nu -\kappa -4\gamma
 ^2+4\sqrt{\gamma ^2+\alpha ^2}
 \vphantom{\sum_{n=1}^{\infty }
 \sum_{I=1}^{4}sign C_I^{(n)}\,\omega _{I,n}-n}
 \right.
 \nonumber \\
 &&\left.
 +\frac{1}{2}\sum_{n=1}^{\infty }
 \sum_{I=1}^{4}\left(sign C_I^{(n)}\,\omega _{I,n}-n-\frac{\kappa
 ^2}{2n}\right)
 \right]\,.
\end{eqnarray}
The last sum can be seen to absolutely converge if we use the
asymptotic values of the frequencies $\omega _{I,n}$ from
\cite{Park:2005ji}. The asymptotic expansion of $F(\beta ,\alpha )$
in $1/\alpha $ terminates at the second order:
\begin{equation}\label{approximation}
 F(\beta ,\alpha )=-\alpha ^2\ln\left(\frac{\,{\rm e}\,^{C-1/2}}{2}\,
 \alpha \right)+\frac{1}{6}+
 O\left(\,{\rm e}\,^{-\alpha }\right),
\end{equation}
where $C=0.5772\ldots $ is the Euler constant.  The dependence on
the fractional part of $\gamma $ is therefore non-perturbative in
$1/\alpha $ and thus in $1/\mathcal{J}$. In particular it will not
be seen in the numerical calculations in sec.~5 which will be done
for sufficiently large values of $\mathcal{J}$.


\subsection{Perturbative expansion}

It is hard to find a useful integral representation for the $\sltwo$
modes because of the sign factors in (\ref{dEosc}). In computing the
perturbative $1/\mathcal{J}$ expansion of the string energy shift we
shall follow a more straightforward approach of evaluating the sum
by first expanding all the frequencies in $1/\mathcal{J}$ and then
computing the sum order by order in $1/\mathcal{J}$. As was already
observed in \cite{Frolov:2004bh} this procedure is not so harmless,
because the sum is not uniformly convergent and modes with $n\sim
\mathcal{J}^2$ can give a finite contribution. This is reflected in
superficial divergences which arise starting from second order in
$1/\mathcal{J}^2$. We shall ignore these problems and will use
zeta-function regularization to sum the divergent series. This
approach might not look well motivated but we shall find a
surprising agreement of this naive summation prescription with the
Bethe ansatz to third order in $1/\mathcal{J}^2$, which gives us a
hint that this prescription may be the correct way to compute the
energy correction on the string theory side.

Using the pertubative expressions for the mode frequencies, which
are given in appendix B,  we can write the pertubative expression
for the energy shift $\delta E$ in powers of $1/\mathcal{J}^2$
\begin{equation}\label{}
 \delta E^{string}=\sum_{p=1}^{\infty }
 \frac{\delta E_p^{string}}{\mathcal{J}^{2p}} \,.
\end{equation}
 It
is given by
\begin{eqnarray}
\label{StringOneLoop}
 \delta E_1^{string} & =&  {1\over 2} m
 (k-m)+{1\over 2} \sum_{n=1}^\infty 2(k-m)m - n^2 + n\sqrt{n^2
      + 4 m(m-k)}  \,,
      \\
\label{StringTwoLoop}  \delta E_2^{string}
    &=&  -{1\over 8}m (k-m) (4k^2 - 11 km + 3 m^2)  \cr
    &&+\sum_{n=1}^\infty  {1\over 8}\big\{-2 ( k - m )  m (
       4 k^2 - 11 k m + 3 m^2 )  + 2 ( 3 k^2 - 10 k m + 5
       m^2)  n^2 + n^4\big\} \cr
    &&- {n (-4 ( k - m )m ( 5k^2 - 15km + 6m^2 )
     + 2( k - 3m )( 3k - 2m) n^2 +  n^4)
     \over  8 \sqrt{n^2 + 4 m(m-k)}}  \,,
     \\
\label{stringThreeLoop} \delta E_3^{string} &=&
 {1\over 16} ( k - m ) m( 8k^4 - 52k^3m + 89k^2m^2 - 42km^3 + 5m^4)\cr
        && +\sum_{n=1}^\infty {1\over 16} \{
                   2 (k-m)m ( 8k^4 - 52k^3m + 89k^2m^2 - 42km^3 + 5m^4) \cr
        && - ( 15k^4 -128 k^3m + 279k^2m^2 - 202km^3 + 44m^4 ) n^2
        \cr
        &&- ( 15k^2 - 38km +
             19m^2 ) n^4 - n^6\} \cr
        && + {1\over 16 (n^2+ 4m (m-k))^{3/2}}
        \cr && \times\big\{
            4( k - m )^2 m^2( 45k^4 - 324k^3m +
            621k^2m^2 - 370km^3 + 60m^4 ) n \cr
        && - 2( k - m ) m( 53k^4 - 481k^3m +
            1083k^2m^2 - 815km^3 + 192m^4 ) n^3 \cr
        && +
            ( 15k^4 - 218k^3m + 603k^2m^2 - 556km^3 +
            164m^4 ) n^5
   \cr   &&+ ( 15k^2 - 44km + 25m^2 )n^7 + n^9 \big\} \,.
\end{eqnarray}
We shall compare this expression to the energy shift calculated using
the Bethe ansatz in the next section.


\section{Bethe ansatz}


\subsection{Classical limit}\label{class}

Classical solutions for the string moving in $AdS_3\times S^1$ are
uniquely specified by the spectral data of the Lax operator. One can
introduce the spectral density $\rho (x)$ defined on a set of
intervals $C_I=(a_I,b_I)$. The spectral density satisfies
 a singular integral equation \cite{Kazakov:2004nh}
\begin{equation}\label{bcl}
 2\pint dy\,\,\frac{\rho (y)}{x-y}=
 2\pi k_I-2\pi \left(\frac{\mathcal{J}+m}{x-1}+
 \frac{\mathcal{J}-m}{x+1}\right),\qquad x\in C_I.
\end{equation}
This can be called the classical Bethe equation, as such type of
equations arise in the thermodynamic limit of quantum Bethe
equations.

In addition, the density obeys a set of normalization conditions
\begin{eqnarray}\label{norm1}
 \int dx\,\,\frac{\rho (x)}{x}&=&-2\pi m,
 \\ \label{norm0}
 \int dx\,\,\frac{\rho (x)}{x^2}&=&
 2\pi (\mathcal{E}-\mathcal{S}-\mathcal{J}),
 \\ \label{norm2}
 \int dx\,\rho (x)
 &=&
 2\pi (\mathcal{E}+\mathcal{S}-\mathcal{J}).
\end{eqnarray}
Here $2\pi m$ is the total world-sheet momentum which must be
quantized because of the periodic boundary conditions on the
world-sheet coordinates.

We shall consider the simplest solutions of (\ref{bcl}) with only
one cut $C=(a,b)$ which corresponds to the circular string
(\ref{classt}). There is only one mode number $k$ in this case. This
simplification is crucial and allows us to rewrite the integral
equation (\ref{bcl}) as an algebraic equation for the resolvant
\begin{equation}\label{}
 G(x)=\int dy\,
 \,\frac{\rho (y)}{x-y}\,.
\end{equation}
The normalization conditions for the density
(\ref{norm1})--(\ref{norm2}) become boundary conditions for $G(x)$
\begin{eqnarray}\label{normal}
 G(0)&=&2\pi m,
 \\ \label{normal1}
 G'(0)&=&-2\pi (\mathcal{E}-\mathcal{S}-\mathcal{J}),
 \\ \label{normal2}
 \lim_{z\rightarrow \infty }zG(z)&=&
 2\pi (\mathcal{E}+\mathcal{S}-\mathcal{J}).
\end{eqnarray}
Multiplying both sides of (\ref{bcl}) by $\rho (x)/(z-x)$ and
integrating over $x$ we find
\begin{equation}\label{bres}
G^2(z)-2\pi \left(k-2\,\frac{\mathcal{J}z+m}{z^2-1}\right)G(z)
 -2\pi \left(\frac{\mathcal{J}+m}{z-1}\,G(1)
 +\frac{\mathcal{J}-m}{z+1}\,G(-1)\right)=0.
\end{equation}
The boundary conditions (\ref{normal})--(\ref{normal2}) can be used
to eliminate $G(\pm 1)$ from this equation. Expanding (\ref{bres})
at $z=0$ and $z=\infty $ we get
\begin{equation}\label{rationality}
 k\mathcal{S}+m\mathcal{J}=0,
\end{equation}
in accord with \cite{Arutyunov:2003za}, and
\begin{equation}\label{gas}
 \left(\mathcal{J}\pm m\right)G(\pm 1)=
 -\pi k(\mathcal{E}+\mathcal{S}-\mathcal{J})\pm \pi m(k+m).
\end{equation}
The condition (\ref{rationality}) imposes rationality on the spins
and requires the integers $k$ and $m$ to have opposite signs. We
shall assume for definiteness that $m>0$ and $k<0$.

Plugging (\ref{gas}) back into (\ref{bres}) we get
\begin{equation}\label{bres1}
 G^2(z)-2\pi \left(k-2\,\frac{\mathcal{J}z+m}{z^2-1}\right)G(z) +
 \frac{4\pi ^2}{z^2-1}\left[k(\mathcal{E}+\mathcal{S}-
 \mathcal{J})z-m(k+m)\right]=0.
\end{equation}
The solution of this quadratic equation is
\begin{equation}\label{resolvant}
 G(z)=\pi \left(k-2\,\frac{\mathcal{J}z+m}{z^2-1}\right)
 +\frac{\pi \sqrt{P(z)}}{z^2-1}\,,
\end{equation}
where
\begin{equation}\label{defP}
 P(z)=k^2z^4
 -4k(\mathcal{E}+\mathcal{S})z^3
 +2(2\mathcal{J}^2+2m^2-k^2)z^2
 +4k(\mathcal{E}-\mathcal{S})z
 +k^2.
\end{equation}
The resolvant determines the density through the discontinuity on
the cut
\begin{equation}\label{denst}
 G(x+i0)-G(x-i0)=2\pi i \rho (x),
 \qquad x\in C,
\end{equation}
and we find
\begin{equation}\label{rho}
 \rho (x)=\frac{\sqrt{-P(x)}}{x^2-1}\,.
\end{equation}

We need one extra condition to express the energy in terms of the
spin and the angular momentum. This condition cannot arise from
equation~(\ref{bres}). Instead one should look more closely at the
structure of the density $\rho (x)$. For general values of the
energy, the angular momentum and the spin, the density is real on
two cuts, whereas we have assumed that the solution has only one
cut. This can be made consistent by requiring that the discriminant
of the quartic polynomial (\ref{defP}) is zero, then $P(z)$ has one
double root (fig.1)
\begin{equation}\label{twoeq}
 P(c)=0,\qquad P'(c)=0\,.
\end{equation}
These two equations determine the dependence of the energy on the
angular momenta, $\mathcal{E}=\mathcal{E}(\mathcal{S},\mathcal{J})$,
in a parametric form
\begin{figure}[t]
\begin{center}
\includegraphics*[width=.4\textwidth]{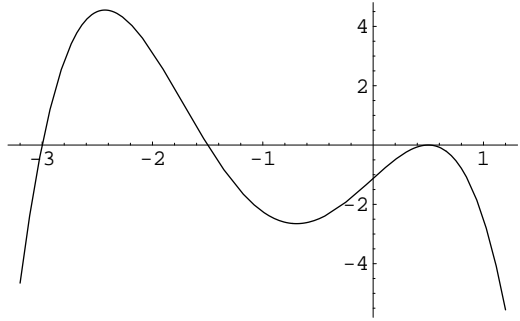}
\end{center}
\caption{Graph of the quartic polynomial $-P(z)$ (the ordering of
the zeroes is $a<b<c$).\label{f:density1}}\end{figure} and are
equivalent to (\ref{ura1}), (\ref{ura2}) upon the identification
\begin{equation}\label{KappaDef}
 \kappa =-\frac{k}{2}\left(\frac{1}{c}-c\right).
\end{equation}


\subsection{Quantum corrections}\label{quantumc}

If the integral equation (\ref{bcl}) is interpreted as the classical
limit of some Bethe equations\footnote{Bethe ansatz only works for
integrable systems, so here we must assume quantum intergrability
of the world-sheet sigma-model. There are indeed some indications
that integrability is not destroyed by quantum corrections
\cite{intr}.}, the density $\rho (x)$ has the
meaning of an asymptotic distribution of Bethe roots in the limit
when their number (naturally identified with the spin $S$ of the
quantum string state) becomes infinite
\begin{equation}\label{nashaplotnost}
 \rho (x)=\frac{4\pi }{\sqrt{\lambda }}
 \sum_{k=1}^{S}\frac{x_k^2}{x_k^2-1}\,\delta (x-x_k).
\end{equation}
The normalization factor $2\pi /\sqrt{\lambda }$ is the coupling
constant of the world-sheet sigma-model. The classical
(weak-coupling) limit corresponds to $\lambda \rightarrow \infty $.
Because $S$ scales with $\sqrt{\lambda }$ according to (\ref{char}),
the classical limit coincides with the thermodynamic limit, in which
the number of roots becomes infinite.

Our starting point are the quantum Bethe equations proposed in
\cite{Staudacher:2004tk,Beisert:2005fw}\footnote{Although the
quantum string can fluctuate in all directions in $AdS_5\times S^5$,
the quantum string Bethe equations have the same number of degrees
of freedom as in the pure $\sltwo$ sector. On the gauge theory side
different sectors do not talk to each other because operators with
different quantum  numbers do not mix \cite{beisert},
but it is not a priori clear
why various sectors can be separated on the string theory side (see
\cite{Minahan:2005jq} for a more detailed discussion of this issue).
}
\begin{equation}\label{bsbethe}
 \left(\frac{x_k^+}{x_k^-}\right)^J
 =\prod_{j\neq k}^{}
 \frac{x_k^--x_j^+}{x_k^+-x_j^-}\,\,
 \frac{1-\frac{1}{x_k^-x_j^+}}{1-\frac{1}{x_k^+x_j^-}}
 \left(\frac{1-\frac{1}{x_k^-x_j^+}}{1-\frac{1}{x_k^+x_j^+}}\,\,
 \frac{1-\frac{1}{x_k^+x_j^-}}{1-\frac{1}{x_k^-x_j^-}}\right)^{
 \frac{i\sqrt{\lambda }(u_k-u_j)}{2\pi }},
\end{equation}
where\footnote{Our notation differs from that of
\cite{Beisert:2005fw} by a rescaling of $x_k$ and $u_k$:
$x_k\rightarrow x_k\sqrt{\lambda }/4\pi $, $u_k\rightarrow
u_k\sqrt{\lambda }/4\pi $.}
\begin{equation}\label{}
 u_k=x_k+\frac{1}{x_k}
\end{equation}
and
\begin{equation}\label{xpm}
 x_k^\pm+\frac{1}{x_k^\pm}=u_k\pm\frac{2\pi i}{\sqrt{\lambda }}\,.
\end{equation}
These equations reduce to (\ref{bcl}) in the thermodynamic limit
when $\sqrt{\lambda },J,S\rightarrow \infty $. Our goal will be to
compute the leading-order quantum correction to the classical Bethe
equations.

It might seem that (\ref{bsbethe}) can only give rise to even powers
of $1/\sqrt{\lambda }$, since the equations are invariant under
$\sqrt{\lambda }\rightarrow -\sqrt{\lambda }$. Nevertheless the odd
powers of $1/\sqrt{\lambda }$ arise in the expansion and the leading
quantum correction is $O(1/\sqrt{\lambda })$ for the following
reason. The Bethe roots $x_k$ condense into cuts in the
thermodynamic limit such that the distance between nearby roots goes
to zero. But the simultaneous limit of $\lambda \rightarrow \infty $
and $x_{k+1}-x_k\rightarrow 0$ is singular in the Bethe equations
and this singularity gives rise to a local anomaly
\cite{kazakovunpublished}. The anomaly cancels at the leading order
\cite{Beisert:2005di}, but contributes to the $1/\sqrt{\lambda }$
quantum correction \cite{Beisert:2005mq,Hernandez:2005nf}. We shall
calculate the anomaly directly from the Bethe equations
(\ref{bsbethe}). The calculations are rather complicated and the
details are given in appendix A. The resulting equation for the
resolvant differs from (\ref{bres1}) by a correction term
\begin{eqnarray}\label{AlgebraicResolv}
 &&G^2(z)-2\pi \left(k-2\,\frac{\mathcal{J}z+m}{z^2-1}\right)G(z)
 +\frac{4\pi ^2}{z^2-1}\left[k(\mathcal{E}+\mathcal{S}
 -\mathcal{J})z
 -m(k+m)
 \vphantom{+\frac{z}{\sqrt{\lambda }}
 \int_{}^{}dx\,\,\frac{x\rho '(x)\coth \pi \rho (x)}{z-x}}
 \right]\nonumber \\
 &&
\qquad  +{4 \pi \over \sqrt{\lambda}}\, {z^2\over z^2-1}
 \int_{}^{}dx\,\,\frac{\rho '(x)\pi \rho (x)
 \coth \pi \rho (x)}{z-x}
 =0.
\end{eqnarray}

Solving this quadratic equation we find a density which is of the
form (\ref{rho}), where the function  $P(z)$ obtains a correction
\begin{equation}\label{}
 \delta P(z)={4 \pi \over \sqrt{\lambda}}\frac{z^2(1-z^2)}{\pi^2 }
 \int_{}^{}dx\,\,\frac{\rho '(x)
 \pi \rho (x)\coth \pi \rho (x)}{z-x}\,.
\end{equation}
The energy can be found as before, from the requirement that there
is only one cut present
\begin{equation}\label{deltaP}
 P(c+\delta c)+\delta P(c+\delta c)=0,\qquad P'(c+\delta c)
 +\delta P'(c+\delta c)=0.
\end{equation}
Expanding the first equation to linear order we get
\begin{equation}\label{}
 \frac{\partial P(c)}{\partial \mathcal{E}}\,\delta \mathcal{E}
 +\frac{\partial P(c)}{\partial c}\,\delta c
 +\delta P(c)=0.
\end{equation}
Taking into account that $\partial P(c)/\partial c=0$ we find
\begin{equation}\label{denergy}
 \delta \mathcal{E}=-
 \frac{\delta P(c)}{\partial P(c)/\partial
 \mathcal{E}}\,.
\end{equation}
For $\partial P/\partial \mathcal{E}$ we get from (\ref{defP})
\begin{equation}\label{dpdc}
 \frac{\partial P(c)}{\partial \mathcal{E}}=-4kc(c^2-1).
\end{equation}
Rescaling back to the physical energy we obtain
\begin{equation}\label{final}
 \delta E^{ Bethe} = {c\over \pi k } \int_{}^{}dx\,\,
 \frac{\rho '(x)\pi \rho (x)\coth \pi \rho (x)}{x-c}\,.
\end{equation}

We can also introduce
\begin{equation}\label{TildeRho}
\tilde{\rho} (x) = {1\over \pi}\int_0^{\pi \rho(x)} d\xi \xi \coth
\xi\,.
\end{equation}
Then integration by parts in (\ref{final}) yields
\begin{equation}\label{IntParts}
\delta E^{ Bethe} = {c\over \pi k } \int dx {\tilde{\rho}(x) \over
(x-c)^2} \,.
\end{equation}

Let us see how the one-loop SYM result
\cite{Beisert:2005mq,Hernandez:2005nf} is recovered. From
(\ref{twoeq}), (\ref{defP}) we find that $c=-k/(2 \mathcal{J})$ at
large $\mathcal{J}$. Inserting this into (\ref{IntParts}) and
rescaling $x\rightarrow 4 \pi \mathcal{J} x$, we get for the energy
shift at the leading order in $1/\mathcal{J}$
\begin{equation}\label{OneLoopE}
\delta E_1^{{ Bethe}} =  -{1\over 8 \pi^2 \mathcal{J}^2} \int
dx\,\, {\tilde{\rho}
  (x)\over x^2} \,,
\end{equation}
in agreement with \cite{Beisert:2005mq}.



To perturbatively evaluate the integral (\ref{final}), we shall need
to expand various parameters characterizing the classical string
configuration in a power series in  $1/{\mathcal J}$. In particular,
we need to find the zeroes of the quartic polynomial $P(x)$.  Recall that
$P(x)$ defined in (\ref{defP}) can be factorized as
\begin{equation}
\label{abc} P(x) =  (x-a) (x-b) (x-c)^2 \,,
\end{equation}
For our sign choice ($m>0$, $k<0$), the roots are ordered as
$a<b<c$.

 The zeroes $a,b,c$  admit an expansion in ${1\over {\cal
J}}$. Solving (\ref{twoeq}) perturbatively in $1/\mathcal{J}$ we get
\begin{eqnarray}\label{cESol}
 c&=&-\frac{k}{2\mathcal{J}}
 +\frac{k}{8\mathcal{J}^3}(2m^2-4 mk + k^2)
 \nonumber \\
 &&+\frac{k}{16\mathcal{J}^5}(-3m^4 + 16m^3k - 23 m^2 k^2 + 10 m k^3 - k^4)
 +O\left(\frac{1}{\mathcal{J}^7}\right),
 \\
 \mathcal{E}&=&\left(1-\frac{m}{k}\right)\mathcal{J}
 +\frac{1}{2\mathcal{J}}\,m(m-k)
 -\frac{1}{8\mathcal{J}^3}\,m(m-k)(m^2-3mk+k^2)
 \nonumber \\
 &&+\frac{1}{16\mathcal{J}^5}\,m(m-k) (m^4 - 7 m^3 k + 13m^2 k^2 - 7 m
 k^3 + k^4)+O\left(\frac{1}{\mathcal{J}^7}\right). \label{Eclass}
\end{eqnarray}
The expression (\ref{Eclass}) agrees with the perturbative expansion
of the classical string energy computed in \cite{Park:2005ji}.




\subsection{Mode expansion}\label{BetheContourSec}

Our starting point is (\ref{final}), which can be written as a
contour integral, because the integrand has a square-root branch cut
along the contour of integration. If we introduce the function
 \be f(z) = {\sqrt{P(z)}\over z^2-1}\,,  \ee
the energy shift becomes
 \be\label{ContourInt} \delta E^{Bethe} =
 {c\over k} \oint_{{\mathcal C}_{ab}} {dz\over 2 \pi i} {f'(z) f(z)
 \cot (\pi f(z)) \over z-c} \,,
 \ee
where the integration contour $\mathcal{C}_{ab}$ encircles the cut clockwise. We can
use the following series representation for $\cot\pi f(z)$
 \be\label{CotExp} \cot (\xi) = {1\over  \xi} + 2 \xi
 \sum_{n=1}^\infty {1\over \xi^2 -n^2 \pi^2 }\,.
 \ee
Inserting this into the contour integral we obtain
 \be\label{BetheEExpandedCot} 
 \delta E^{Bethe} =
             {c\over k} \oint_{\mathcal{C}_{ab}} dz
                   {f'(z)\over (z-c)}
           + {2c\over k} \sum_{n =1}^\infty
                       \oint_{\mathcal{C}_{ab}} dz   { f'(z) f^2(z) \over (z-c)
             (f^2(z) -n^2)} \,.
 \ee
The only singularities of the integrands outside the contour of
integration are poles and  the integrals can be calculated by
evaluating the residues. The integrand in the first term has poles
at $z=c$ and $z=\pm 1$. The poles of the second term are at $z=\pm
1$ and at $z=z_n$, where the $z_n$'s are solutions of
\be\label{fzn}
f(z_n)=\pm n \,,\qquad n\in\mathbb{N} \,.
\ee
Squaring this equation we find that $z_n$'s are
the roots of the quartic equation
\begin{equation}\label{pofz}
 P(z)=n^2(z^2-1)^2.
\end{equation}
It can be shown that the fluctuation energies around the classical
solution are determined by the same equation, in accord with the
general relationship between fluctuations \cite{Beisert:2003xu} and
finite-size corrections for Bethe ansatz \cite{Beisert:2005bv}. The
residues at $z=\pm 1$ are rather complicated, but the residues at
$z=z_n$ are easy to evaluate
\be\label{Reszn}
 \hbox{Res}_{z=z_n}    ={c\over k} \left({n \epsilon_n \over
                           z_n-c}  \right).
\ee
The sign $\epsilon_n$ of the residue is the same as the sign in the equation
$f(z_n)=\pm n$ and can be determined by analyzing (\ref{pofz})
with the help of (\ref{abc})
 \be\label{EpSign} \epsilon_n =\left\{ \ba +1
 &\quad \hbox{for}\quad  z\in [-\infty, a] \cup [-1, c]\cup [1,
 \infty] \cr -1 &\quad \hbox{for}\quad  z\in [b, -1]\cup [c, 1] \,.\ea
 \right. \ee


\subsection{Perturbative expansion and comparison to string theory}

We have  evaluated the residues in (\ref{BetheEExpandedCot})
perturbatively in $1/\mathcal{J}$. The calculations are lengthy and
are given in appendix C. We also checked that the
first two orders are reproduced by a direct expansion of the
integral (\ref{IntParts}). Unlike the string sum over modes, its
Bethe counterpart is manifestly finite at each order of the
perturbative expansion. This might indicate that our method of
computing the series over string modes breaks down at two loops (see
also the discussion in \cite{Frolov:2004bh}). However, if we compare
the zeta-regularized sum (\ref{StringOneLoop}),
(\ref{StringTwoLoop}) and (\ref{stringThreeLoop}) with the Bethe
ansatz, we find complete agreement! We checked this up to the third
order
\begin{equation}\label{}
 \delta E^{Bethe}_p=\delta E^{string}_p,\qquad p=1,2,3.
\end{equation}
The agreement at the first two orders implies that the string energy
shifts agree with the finite-size corrections to the anomalous
dimensions at two loops in the SYM theory. At three loops, the
string Bethe ansatz that was our starting point, differs from the
gauge Bethe ansatz \cite{Beisert:2004hm} which computes the
anomalous dimensions.

The agreement between the Bethe ansatz and the direct string
calculation is rather spectacular. The initial expressions look too
complicated for this to be a pure accident. Nevertheless, the string
and the Bethe calculation have a different status. The Bethe ansatz
energy shift is automatically finite order by order in
$1/\mathcal{J}$. On the string side we encountered divergences
despite the complete, unexpanded  energy shift being finite. No
doubt, there should be a better way to approach the weak-coupling
(large $\mathcal{J}$) limit on the string side.

\section{Limit of large winding number}
\label{sff}

Because of the divergences in the naive $1/\mathcal{J}$ expansion of
the string sum, it would be desirable to do an independent test
which avoids the convergence issues mentioned earlier. One option is
to evaluate the energy shifts numerically. This is done in the next
section. Here we consider  a particular regime, the limit of large
winding number ($|k|\gg 1 $), in which the energy shifts can be
calculated analytically\footnote{In the narrow sense, we are just
comparing two mathematical expressions -- the string one-loop
corrections (\ref{full})--(\ref{StringEnergyShift}) and the
finite-size correction from the Bethe ansatz (\ref{final}). Each is
a well-defined function of the parameters $k$, $m$ and
$\mathcal{J}$. If the two expressions agree (or disagree), they must
agree (disagree) at all values of the parameters, in particular if
one of the parameters ($k$ in this case) takes its extreme value.
From this point of view the limit of large $k$ is just a simplifying
assumption that allows us to calculate $\delta E^{String}$ and
$\delta E^{Bethe}$ explicitly in some corner of the parameter space.
On the other hand, not only the classical energy of the string, but
also the quantum correction to it stays finite in the large-$k$
limit. This probably means that the limit of large winding (or small
spin) is well-defined for this type of string solutions and it would
be very interesting to study this limit further. The winding number
in that, more general setting should be much larger than the
rescaled quantities $\mathcal{E}$ and $\mathcal{J}$, but should be
much smaller than $\sqrt{\lambda }$ (and thus $E$ and $J$) in order
not to interfere with the loop expansion of the sigma-model.}. In
this limit $\mathcal{J}$, $\mathcal{E}$ and $m$ stay finite, but the
spin goes to zero: $\mathcal{S}\ll 1$. The string remains
macroscopic in this limit, since it winds the big circle of $S^5$,
but in $AdS_5$ the string shrinks to zero size ({\it cf.}
(\ref{ura})). We will have to assume that $\mathcal{J}/|k|\ll 1$,
which means that there is no overlap with the perturbative regime we
have discussed so far. In fact, the energy shift turns out to depend
on $1/\mathcal{J}=\sqrt{\lambda}/J$ rather than $1/\mathcal{J}^2$ in
the large-$k$ limit, and it is not possible to compare string
quantum corrections to perturbative SYM theory in this regime.

 The details of the string calculation
 are given in appendix D. The result is
\begin{eqnarray}\label{stringk}
 \delta E&=&\frac{
 2F\left(0,\sqrt{\mathcal{J}^2-m^2}\right)
 +2F\left(0,\mathcal{J}+m\right)
 -4F\left(\{\frac{|k|}{2}\},\sqrt{\mathcal{J}
 (\mathcal{J}+m)}\right)}{\mathcal{J}+m}
 \nonumber \\
 &&+\sqrt{m\mathcal{J}}+(\mathcal{J}+m)\ln\frac{\sqrt{\mathcal{J}+m}}
 {\sqrt{\mathcal{J}}+\sqrt{m}}\, - m \, ,
\end{eqnarray}
where the function $F(\beta ,\alpha )$ is defined in
(\ref{defFbetaalpha}). A peculiar property of this result is the
dependence on the fractional part of $k/2$, which means that the
large-$k$ limit of the string energy shift depends on whether the
winding number $k$ is even or odd. This effect probably arises
because of the $k$-dependent field redefinition of the world-sheet
fermions which was used to find the spectrum of fluctuations
\cite{Frolov:2003tu,Frolov:2004bh,Park:2005ji}. This kind of
irregularity does not arise in the Bethe ansatz, and also in the
zeta-regularized large-$\mathcal{J}$ expansion.


\subsection{Bethe ansatz calculation}

We begin with the classical limit. To take the large-$k$ limit it is
convenient to rewrite (\ref{defP}) in the two equivalent forms
\begin{equation}\label{Plargek}
 P(x)=k^2(x^2-1)^2-4k\mathcal{E}x(x^2-1)+4m\mathcal{J}x(x\pm 1)^2
 +4(\mathcal{J}\mp m)^2x^2.
\end{equation}
The first two terms blow up in the  $k\rightarrow \infty $ limit
unless $x$ is close to $1$ or $-1$. The roots of $P$, $a$, $b$ and
$c$, thus lie in the vicinity of $\pm 1$. Changing the variables to
\begin{equation}\label{}
 x=\pm 1+\frac{v }{k}\,,
\end{equation}
and taking the limit $k\rightarrow \infty $, we get
\begin{equation}\label{newp}
 P(x)=4v ^2-8\mathcal{E}v +4(\mathcal{J}\pm m)^2,
 \qquad {\rm at}\qquad x\rightarrow \pm 1.
\end{equation}
Thus two of the roots of $P(x)$ lie near $1$ and two lie near $-1$.
The double root should lie at $x\approx 1$, from which we find
\begin{equation}\label{}
 \mathcal{E}=\mathcal{J}+m
\end{equation}
and
\begin{equation}\label{}
 c=1-\frac{\mathcal{E}}{|k|}\,.
\end{equation}
Solving (\ref{newp}) near $x=-1$, we find the endpoints of the cut
\begin{equation}\label{}
\left\{  \begin{array}{c}
  b   \\
  a  \\
\end{array}  \right\}
=-1-\frac{\left(\sqrt{\mathcal{J}} \pm\sqrt{m}\right)^2}{|k|}\,.
\end{equation}
We see that the cut shrinks to a very small size, whereas the
density according to (\ref{norm1})-(\ref{norm2}) is still normalized
to $O(1)$. Thus the density is highly peaked near $-1$. Indeed, from
(\ref{rho}) and (\ref{newp}) we find
\begin{equation}\label{newrrho}
 \rho (x)=\frac{|k|}{v }\,\sqrt{2\left(\mathcal{J}+m\right)v
 -v ^2-\left(\mathcal{J}-m\right)^2 }\,.
\end{equation}

The integral (\ref{IntParts}) can be easily evaluated in the
$k\rightarrow \infty $ limit. Because the density is large, $\cosh
\xi $ in (\ref{TildeRho}) can be approximated by $1$, and thus
\begin{equation}\label{}
 \tilde{\rho }=\frac{\pi }{2}\,\rho ^2, \qquad
 {\rm at}\qquad \rho \rightarrow \infty .
\end{equation}
We thus get from (\ref{IntParts})
\begin{equation}\label{}
 \delta E^{Bethe}=\frac{1}{8k}\int_{}^{}dx\,\rho ^2(x).
\end{equation}
Using $dx=dv /|k|$ and the explicit expression (\ref{newrrho}) for
the density, we find
\begin{equation}\label{}
 \delta E^{Bethe}=\sqrt{m\mathcal{J}}-\frac{\mathcal{J}+m}{2}\,\ln
 \frac{\sqrt{\mathcal{J}}+\sqrt{m}}{\sqrt{\mathcal{J}}-\sqrt{m}}
 \,.
\end{equation}
This clearly disagrees with the string theory calculation
(\ref{stringk}), in particular the Bethe ansatz result has a regular
dependence on $k$. We shall see this discrepancy also in the
numerical calculations. Let us also note that
even though the explicit computation in this
section was done in the simplifying large~$k$ limit, the deviations
between the Bethe ansatz and the string theory computation are also
observed numerically for finite values of the parameter~$k$
(see figure~\ref{f:fig1} in the next section).



\section{Numerical evaluation of energy shifts}
\label{numer}

In this section we numerically compare corrections to
the energy of the circular string obtained by the semiclassical
quantization (\ref{StringEnergyShift}) and the one deduced from the
proposed quantum string Bethe equation (\ref{final}). Both
evaluations of the sums are done for various values of the
parameters.
We first consider the large-$\mathcal{J}$ limit.
From figure 2 we see that both functions have the same leading order behaviour, in agreement with the earlier analytic
results.
Next, we try to extract
the coefficients of the $1/\mathcal{J}^2$ expansion of the energy
shift numerically. In practice, numerically computing higher order effects is hard, since it requires a high numerical precision and
stability.

\begin{figure}[t]
\begin{center}
\includegraphics*[width=.5\textwidth]{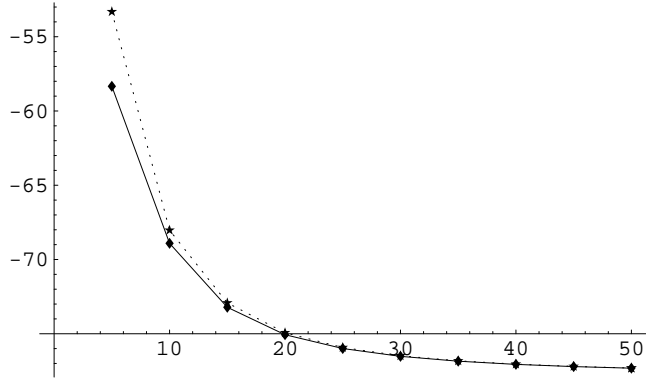}
\end{center}
\caption{Energy shifts $(\delta E)\times {\mathcal J}^2$
for ${\mathcal J} = 5...50 \, ,  \, m=3\,  ,  \, k= -2  $, Bethe
vs. semi-classical string.}\label{f:LowJ}
\end{figure}

Yet, by using high precision numerical evaluations
let us  try to extract the first subleading ($1/{\mathcal J^2}$)
correction from the exact semiclassical expression
(\ref{StringEnergyShift}) and compare it with the zeta-function
regularized result (\ref{StringTwoLoop}). Subtracting the analytic
one-loop piece (\ref{StringOneLoop}) from the numerical expression
for the semiclassical energy shift (\ref{StringEnergyShift}) leads
to very unstable numerical results, given in table \ref{table1}.
\begin{equation}
\begin{gathered}
$m=3.0\, , \quad \quad k=-2$ \,.\\
\hbox{\begin{tabular}{|r|c|c|c|c|c|c|c|c|c|l|}
\hline
\label{table1}
${\mathcal J}  \quad \quad \quad \quad $ & 50& 100 & 150 & 200& 250  \\
\hline
 $(\delta E^{string}- \delta E_1) \times {\mathcal J}^2 $  &
 1041 &620  & -82  & -1066  & -2329  \\
\hline
$ {\mathcal J} \quad \quad \quad \quad   $  &300 & 350 &  400 & 450 & 500 \\
\hline
 $(\delta E^{string}- \delta E_1) \times {\mathcal J}^2 $  & -3871 & -5693 & - 7794 & -10174 & - 12831 \\
\hline
\end{tabular}}
\end{gathered}
\end{equation}
This should be compared to the zeta-function regularized two-loop
result (\ref{StringTwoLoop}) for the same values of $m$ and $k$
which gives
\begin{equation}
\label{2loop}
\delta E_2 = 393.375\,.
\end{equation}
The numerical stability is greatly improved, if instead of
subtracting the analytic one-loop result (\ref{StringOneLoop}), we
use the asymptotic numerical value for the energy shift (obtained
for ${\mathcal J}=10^3$)
\begin{equation}
\delta E^{string}_{asymptot} = -77.781\,.
\end{equation}
The results are given in table (\ref{table2}). We see that it is
much less fluctuating compared to the result in table
(\ref{table1}). The deviations from the constant value, may be
attributed to higher orders in $1/{\mathcal J}^2$ and insufficient
numerical precision. The average value from the table (\ref{table2})
is different from the regularized two-loop result (\ref{2loop}), but
the numerics is rather unstable and we cannot draw any definite
conclusions at this point because of insufficient numerical
accuracy.
\begin{equation}
\begin{gathered}
$m=3.0\, , \quad \quad k=-2$ \,,\\
\hbox{\begin{tabular}{|r|c|c|c|c|c|c|c|c|c|l|}
\hline
\label{table2}
${\mathcal J}  \quad \quad \quad \quad $ & 50& 100 & 150 & 200& 250  \\
\hline
 $(\delta E^{string}- \delta E^{asymptot}) \times {\mathcal J}^2 $  &
 1170 &1167  & 1147  & 1120  & 1087  \\
\hline
${\mathcal J}\quad \quad \quad \quad $  &300 & 350 &  400 & 450 & 500 \\
\hline
 $(\delta E^{string}- \delta E^{asymptot}) \times {\mathcal J}^2 $  & 1048 & 1004 & 952 & 896 & 835 \\
\hline
\end{tabular}}
\end{gathered}
\end{equation}

\begin{figure}[t]
\begin{center}
\includegraphics*[width=.5\textwidth]{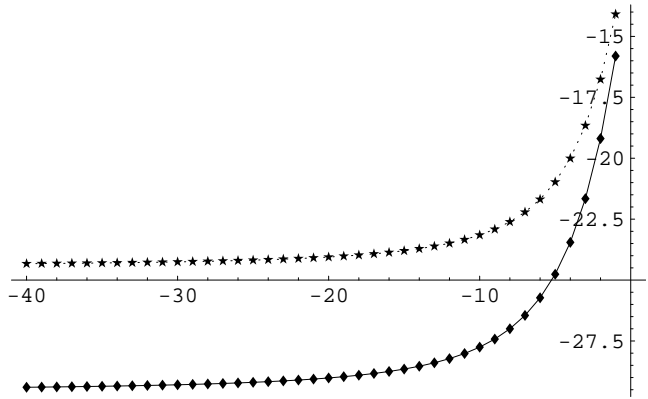}
\end{center}
\caption{Energy shifts $(\delta E)\times {\mathcal J}^2$ for
${\mathcal J} = 3 \, ,  \, m=2\, , \,  -k=(40...1)$. The upper curve
is the string calculation. The lower curve is the prediction of the
Bethe ansatz.}\label{f:fig1}
\end{figure}

We get much better accuracy if we look at a finite value of
${\mathcal J}$ and vary $k$ at fixed $m$ and $\mathcal{J}$.
We shall take ${\mathcal J}=3$ and $m=2$ and vary $k$ from $-40$ to
$-1$. The results are given in figure \ref{f:fig1}. The upper curve
is the semiclassical string computation, the lower curve is computed
from the Bethe ansatz\footnote{By that we mean numerical integration
in (\ref{final}). Direct numerical solution of the discrete Bethe
equations with subsequent extrapolation to the thermodynamic limit
requires substantially more involved calculations.}. We see that
both the semiclassical and the Bethe energy shifts tend
asymptotically to constant but different values, which are in good
numerical agreement with the analytic calculations in the previous
section. Here our numerical precision is sufficient to discriminate
the two results.

\section{Conclusions}

We have compared quantum correction to the energy of macroscopic
rigid strings in $AdS_5\times S^5$ with the finite-size corrections
to the quantum string Bethe ansatz. Taken at face value, the two
results disagree, but an interpretation of this discrepancy is
unclear to us. If we do the string calculation in a more naive way
by first expanding fluctuation frequencies in $1/\mathcal{J}$ and
then summing the series over string modes, the straightforward
zeta-regularized expansion in $1/\mathcal{J}^2$ agrees with the
Bethe ansatz to the first three orders. Perhaps the sum over
frequencies on the string side should be redefined such that it
automatically reproduces zeta-regularized $1/\mathcal{J}$ expansion.
The methods used to evaluate related sums in the context of
plane-wave string theory \cite{Lucietti:2003ki} can be helpful to
implement such zeta-function prescription. On the other hand the sum
is finite and well-defined as it stands and there are no apparent
regularization ambiguities.

Another possible explanation of the discrepancy is that the string
Bethe equations receive non-trivial $1/\sqrt{\lambda }$ corrections.
We cannot discriminate between these two possibilities at present.
Studying other classes of string solutions will be certainly helpful
to resolve this puzzle. We should first of all mention stable
circular strings on $S^5$ which were analyzed both in string theory
\cite{Frolov:2003tu} and using the Bethe ansatz
\cite{Engquist:2003rn}.


\subsection*{Acknowledgements}

We would like to thank G.~Arutyunov, N.~Beisert, V.~Kazakov, J.~Lucietti,
J.~Minahan, M.~Staudacher and A.~Tseytlin for interesting
discussions. The work of S.S.-N. was partially supported by
the DFG, DAAD, and European RTN Program MRTN-CT-2004-503369. The
work of K.Z. was supported in part by the Swedish Research Council
under contracts 621-2002-3920 and 621-2004-3178, and by the G\"oran
Gustafsson Foundation.



\setcounter{section}{0}
\appendix{Calculation of anomaly}

In this appendix the anomaly term is derived from the quantum string
Bethe equations (\ref{bsbethe}). The following integral
representation turns out to be useful
\begin{equation}\label{}
 \ln\frac{f(x_1^+,\ldots ,x_S^+;x_1^-,\ldots ,x_S^-)}
 {f(x_1,\ldots ,x_S;x_1,\ldots ,x_S)}
 =i\int_{0}^{\frac{2\pi}{\sqrt{\lambda }}}d\varepsilon \,
 \,\frac{1}{f}\sum_{k=1}^{S}\left(
 \frac{x_k^{+\,2}}{x_k^{+\,2}-1}\,\,\frac{\partial f}{\partial x_k^+}
 -\frac{x_k^{-\,2}}{x_k^{-\,2}-1}\,\,\frac{\partial f}{\partial x_k^-}
 \right),
\end{equation}
where $f$ is an arbitrary function and
\begin{equation}\label{}
 x_k^\pm+\frac{1}{x_k^\pm}=u_k\pm i\varepsilon\,,
\end{equation}
under the integral ($x_k^\pm$ on the left-hand-side is defined in
(\ref{xpm})). This representation singles out a particular branch of
the logarithm, so when we write the Bethe equations (\ref{bsbethe})
in the logarithmic form, we should introduce an arbitrary phase
which parameterizes different branches of the logarithm
\begin{eqnarray}\label{}
 &&\int_{0}^{\frac{2\pi }{\sqrt{\lambda }}}d\varepsilon \,
 \left\{
 J\left(\frac{x_k^{+}}{x_k^{+\,2}-1}+\frac{x_k^{-}}{x_k^{-\,2}-1}
 \right)
 +\sum_{j\neq k}^{}
 \left[
 \left(\frac{x_k^{+\,2}}{x_k^{+\,2}-1}+\frac{x_j^{-\,2}}{x_j^{-\,2}-1}
 \right)\frac{1}{x_k^+-x_j^-}+
 \right.\right.\nonumber \\ &&\left.\left.
 \left(\frac{x_k^{-\,2}}{x_k^{-\,2}-1}+\frac{x_j^{+\,2}}{x_j^{+\,2}-1}
 \right)\frac{1}{x_k^--x_j^+}
 \right]
 -\sum_{j\neq k}^{}
 \left[
 \frac{x_j^+}{(x_j^{+\,2}-1)(x_k^-x_j^+-1)}
 +\frac{x_j^-}{(x_j^{-\,2}-1)(x_k^+x_j^--1)}
 \right.\right. \nonumber \\ &&\left.\left.
 -\frac{x_k^+}{(x_k^{+\,2}-1)(x_k^+x_j^--1)}
 -\frac{x_k^-}{(x_k^{-\,2}-1)(x_k^-x_j^+-1)}
 \right]
 \right. \nonumber \\ &&\left.
 -\frac{i\sqrt{\lambda }}{2\pi }\sum_{j\neq k}^{}(u_k-u_j)
 \left[
 \frac{x_k^{+\,2}(x_j^+-x_j^-)}
 {(x_k^{+\,2}-1)(x_k^+x_j^--1)(x_k^+x_j^+-1)}
 +\frac{x_k^{-\,2}(x_j^+-x_j^-)}
 {(x_k^{-\,2}-1)(x_k^-x_j^+-1)(x_k^-x_j^--1)}
 \right.\right. \nonumber \\ &&\left.\left.
 +\frac{x_j^{+\,2}(x_k^+-x_k^-)}
 {(x_j^{+\,2}-1)(x_k^+x_j^+-1)(x_k^-x_j^+-1)}
 +\frac{x_j^{-\,2}(x_k^+-x_k^-)}
 {(x_j^{-\,2}-1)(x_k^+x_j^--1)(x_k^-x_j^--1)}
 \right]
 \right\}
 =2\pi k.
\end{eqnarray}
An important property of this terrible-looking equation is the
symmetry with respect to $\varepsilon \rightarrow -\varepsilon $,
which means that the direct strong-coupling expansion starts from
order $O(1/\lambda )$. The only source of $1/\sqrt{\lambda }$
corrections is the first sum over $j$, in which terms with $j\sim k$
become singular in the $\varepsilon \rightarrow 0$ limit. The
contribution of these terms is the anomaly. In the remaining terms
we can take the limit $\varepsilon \rightarrow 0$ directly
\begin{eqnarray}\label{bbb}
 &&\frac{4\pi \mathcal{J}x_k}{x_k^2-1}+\frac{4\pi }{\sqrt{\lambda }}
 \sum_{jk}^{}\frac{x_k-x_j}{(x_k^2-1)(x_j^2-1)}+
 \int_{0}^{\frac{2\pi }{\sqrt{\lambda }}}d\varepsilon \,
 \sum_{j\neq k}^{}
 \left[
 \left(\frac{x_k^{+\,2}}{x_k^{+\,2}-1}+\frac{x_j^{-\,2}}{x_j^{-\,2}-1}
 \right)\frac{1}{x_k^+-x_j^-}+
 \right.\nonumber \\ &&\left.
 \left(\frac{x_k^{-\,2}}{x_k^{-\,2}-1}+\frac{x_j^{+\,2}}{x_j^{+\,2}-1}
 \right)\frac{1}{x_k^--x_j^+}
 \right]=2\pi k,
\end{eqnarray}
where we have used the equality
$$
u_k-u_j=\frac{(x_k-x_j)(x_kx_j-1)}{x_kx_j}\,.
$$

The next step is to multiply both sides of (\ref{bbb}) by
$1/(z-x_k)$ and sum over $k$. Because of the anti-symmetry in $k$
and $j$, in the double sums $1/(z-x_k)$ can be replaced by
$$
\frac{1}{z-x_k}\rightarrow \frac{1}{2}\left(
\frac{1}{z-x_k}-\frac{1}{z-x_j}\right)
=\frac{x_k-x_j}{2(z-x_k)(z-x_j)}\,.
$$
Now we can disentangle the ``normal" contribution of $j-k\sim \sqrt{
\lambda }$ from the local ``anomalous" contribution of
$j-k\ll\sqrt{\lambda }$. In the latter case
\begin{equation}\label{}
 x_{j+n}\approx x_j+
 \frac{4\pi x_j^2n}{\sqrt{\lambda }(x_j^2-1)\rho (x_j)}\,,
\end{equation}
according to the definition of the density in (\ref{nashaplotnost}).
Also,
\begin{equation}\label{}
x_{j+n}^\pm\approx x_j+\frac{x_j^2}{x_j^2-1}\left(
 \frac{4\pi n}{\sqrt{\lambda }\rho (x_j)}
 \pm i\varepsilon
\right)
\end{equation}
and
$$
\frac{x_{j+n}-x_j}{x_{j+n}^\pm-x_j^\mp}-1 =\mp\frac{2i\varepsilon
}{\frac{4\pi n}{\sqrt{\lambda }\rho (x_j)}\pm 2i\varepsilon }\,.
$$
Separating the long-distance   contributions from the short-distance
ones we find, after some calculations
\begin{eqnarray}\label{bbbe}
&&G^2(z)-2\pi \left(k-2\,\frac{\mathcal{J}z+m}{z^2-1}\right)G(z) +
 \frac{4\pi ^2}{z^2-1}\left[k(\mathcal{E}-\mathcal{S}-
 \mathcal{J})z-2m\mathcal{J}z-m(k+m)\right]\
 \nonumber \\ &&-\frac{4\pi }{\sqrt{\lambda }}\,\,
 \frac{z^2}{z^2-1}
 \sum_{j}^{}\frac{2x_j^2}{(x_j^2-1)(z-x_j)^2}
 \int_{0}^{\frac{2\pi }{\sqrt{\lambda }}}d\varepsilon \,
 \sum_{n=-\infty }^{n=+\infty }\frac{\varepsilon ^2}
 {\frac{4\pi ^2n^2}{\lambda \rho ^2(x_j)}+\varepsilon ^2} =0,
\end{eqnarray}
where
\begin{equation}\label{}
 G(z)=\frac{4\pi }{\sqrt{\lambda }}\sum_{k}^{}
 \frac{x_k^2}{x_k^2-1}\,\,\frac{1}{z-x_k}\,.
\end{equation}

The asymptotics of (\ref{bbbe}) at $z\rightarrow \infty $ shows that
the condition (\ref{rationality}) does not receive quantum
corrections. Performing the summation in the anomaly term  and
changing the integration variable to $\xi =\sqrt{\lambda
}\rho\varepsilon /2$ we finally get
\begin{eqnarray}\label{}
 && G^2(z)-2\pi \left(k-2\,\frac{\mathcal{J}z+m}{z^2-1}\right)G(z) +
 \frac{4\pi ^2}{z^2-1}\left[k(\mathcal{E}+\mathcal{S}-
 \mathcal{J})z-m(k+m)\right]\nonumber \\
 &&-\frac{4\pi }{\sqrt{\lambda }}\,\,
 \frac{z^2}{z^2-1}\int_{}^{}dx\,\,\frac{\tilde{\rho }(x)}{z-x}=0,
\end{eqnarray}
where $\tilde{\rho }(x)$ is defined in (\ref{TildeRho}). The form of
the anomaly used in the main text is obtained after integrating by
parts and taking into account that
\be
\tilde{\rho }'=\rho '\pi \rho \coth\pi \rho \,.
\ee

\setcounter{subsection}{0}
\appendix{Details of string theory computation}
\label{StringDetails}

\subsection{Contribution of $\sltwo$ modes}

The main difficulty in evaluating the energy from the string theory is
the sum over the roots of the quartic polynomial (\ref{PolyPTT})
\be
\delta E^{\sltwo} = \sum_I sign (C_{I}^{(n)}) \omega_{I, n}\,.
\ee
The quartic equation is equivalently given by
\begin{equation}
\label{quartic1}
 \omega^4 + a_2 \omega^2 + a_1 \omega + a_0  =0 \,,
\end{equation}
where
\begin{equation}
\begin{aligned}
a_2&= - 4 k^2 - 2 n^2 - 4 k^2 r_1^2 - 4 \kappa^2 \cr a_1&= 8 k n
\sqrt{k^2 + \kappa^2} (1 + r_1^2) \cr a_0&= n^4 - 4 k^2 n^2 (1 +
r_1^2) \, .
\end{aligned}
\end{equation}
In particular, the absence of the cubic term implies $\sum_{I=1}^4
\omega_{I, n}=0$. The roots can be written as
\begin{equation}
\begin{aligned}
\omega_{1/2,n}&= {1\over 2} \left( R_n\pm D_n \right) \cr
\omega_{3/4,n}&= {1\over 2} \left( -R_n \pm F_n \right) \,,
\end{aligned}
\end{equation}
where
\begin{equation}\label{DefRDF}
\begin{aligned}
R_n&= \sqrt{ y_1 - a_2} \cr D_n&= \sqrt{-R_n^2 - 2 a_2 - {2 a_1\over
R_n} } \cr F_n&= \sqrt{-R_n^2 - 2 a_2 + {2 a_1\over R_n} } \,,
\end{aligned}
\end{equation}
and $y_1$ is a real root of the discriminant cubic equation
\begin{equation}
y^3 - a_2 y^2 -4 a_0 y + 4 a_2 a_0 - a_1^2 =0\,.
\end{equation}
That is
\begin{equation}
y_1 = {1\over 3} a_2 + \left(M + \sqrt{M^2 + S^3} \right)^{1/3}+
\left(M - \sqrt{M^2 + S^3} \right)^{1/3} \,,
\end{equation}
where
\begin{equation}
\begin{aligned}
S& = {1\over 9} (-12 a_0 - a_2^2)\cr M& = {1\over 54} (27 a_1^2 - 72
a_0 a_2 + 2 a_2^3)  \,.
\end{aligned}
\end{equation}

Furthermore, we need to address the issue of the signs in front of
the frequencies. If we take all square roots with positive sign, it
is clear that for a generic $n$ and ${\mathcal J}$ there are two
possibilities for the relative ordering of the frequencies $\omega_I$
\begin{eqnarray}
\label{order}
&\text{I}:& \omega_4 < \omega_3 < \omega_2 < \omega_1 \, \\
&\text{II}:& \omega_4 < \omega_2 < \omega_3 < \omega_1 \, .
\end{eqnarray}
In order to discriminate these,
consider the large $\mathcal{J} \gg n$ limit. The asymptotics are
$\omega_1 \sim - \omega_4 \sim  2 {\mathcal J}$ and so $(\omega^2 - n^2) > 0$.
Hence,
\begin{equation}
\label{sgn1} sign (C_{1, B}^{(n)}) = +1 \,,\qquad sign (C_{4,
B}^{(n)}) = -1 \,.
\end{equation}
On the other hand, in the same limit we have $\omega_2\sim -\omega_3 \sim
n/2 \mathcal{J}$ and thus $(\omega^2 - n^2) <0$, wherefore
\begin{equation}
\label{sgn2} sign (C_{2, B}^{(n)}) = -1 \,,\qquad sign (C_{3,
B}^{(n)}) = +1 \, .
\end{equation}
Hence, in the large $\mathcal{J}$ limit the eigenvalues are ordered
as in the first case in (\ref{order}).
Note that the ordering of $\omega_n^I$ as a function of $n$ keeping
$\mathcal{J}$ fixed does not change,
\ie, the roots do not ``cross'' (see figure \ref{f:omega}).

Using (\ref{sgn1}) and (\ref{sgn2}) the expression for
$\delta E^{\sltwo}$ in the large ${\mathcal J}$ limit can be
simplified to
\be\label{SLTwoSum}
\delta E^{\sltwo} = \sum_n (-\omega_4 + \omega_3 - \omega_2 +
\omega_1) = 2 \sum_n (\omega_1 + \omega_3) = \sum_n D_n + F_n \,.
\ee
In summary, to compute $\delta E^{\sltwo}$ one only needs to
determine the sum over the combination $D_n+F_n$.


\subsection{Perturbative expansion of modes}

The combination of $\sltwo$ modes, $D_n+F_n$,
has the following expansion in $1/\mathcal{J}$
\be\label{EnergySlTwo} \ba {\delta
E^{\sltwo} \over 2\kappa} &= \sum_n
     \left({2k( k - m )  + n^2 + n \sqrt{4m( m-k)  + n^2}\over 2 }
     \right)  {1 \over \mathcal{J}^2}\cr
&+  \bigg(- {-4m ( k - m ) (5k^2- 15km + 6m^2 )  + 2( k - 3m )(3k - 2m)n^3
            + n^5  \over 8 \sqrt{n^2 + 4 m(m-k)}} \cr
&+   {1\over 8 }
     ( -2 k ( k - m ) (k^2 -11k m + 6 m^2)
       - 2 ( 3 k^2 - 10 k m + 5 m^2)  n^2 - n^4
     )\bigg) {1 \over \mathcal{J}^4} \cr
& + \bigg( {1\over 16} \big\{
         2k( k - m ) ( k^4 - 23k^3m +
     86k^2m^2 - 71km^3 + 15m^4)  \cr
&+
         ( 15k^4 - 128k^3m + 279k^2m^2 - 202km^3 +
     44m^4 ) n^2 +(15k^2 - 38km + 19m^2) n^4 + n^6  \big\}\cr
& + {1\over 16 (n^2+ 4m (m-k))^{3/2}}\big\{
            + 4( k - m )^2 m^2( 45k^4 - 324k^3m +
            621k^2m^2 - 370km^3 + 60m^4 ) n \cr
& -         2( k - m ) m( 53k^4 - 481k^3m +
            1083k^2m^2 - 815km^3 + 192m^4 ) n^3 \cr
& +         ( 15k^4 - 218k^3m + 603k^2m^2 - 556km^3 +
            164m^4 ) n^5 + ( 15k^2 - 44km + 25m^2 )n^7 + n^9
        \big\}\bigg){1 \over \mathcal{J}^6}  \,.
\ea
\ee
The other terms, \ie, the transverse and fermionic terms, are as
follows
\be
\ba
&\delta E-{\delta E^{\sltwo} \over 2\kappa}\cr
&= \sum_n
\left(-( k - m )^2 - n^2\right)  {1 \over \mathcal{J}^2}\cr
& +{1\over 16 } \big((k - m)^2(k^2 - 42km - 7m^2)
        + 8( 3k^2 - 10km + 5m^2) n^2 + 4n^4
  \big) {1 \over \mathcal{J}^4}\cr
& + {1\over 128}\big(
  -(k-m)^2(k^4 - 232k^3m +962k^2m^2 - 80km^3 - 11m^4) \cr
&  -4(15k^4-260k^3m+594k^2m^2-340km^3+23m^4)n^2
  - 16(15k^2-38km+19m^2)n^4 - 16n^6
\big)  {1 \over \mathcal{J}^6}\,.
\ea
\ee
Note the three-loop term, where the expression
at order $n^2$ has a different structure from the one in
(\ref{EnergySlTwo}).

Furthermore, expanding the zero mode part of the energy shift  (\ref{ZeroModeE})
in $1/\mathcal{J}$ we obtain
\be\label{ZeroModeEExpanded}
\ba E^{(0)} &= {1\over 2} m (k-m) {1 \over \mathcal{J}^2} - {1\over
32} (3k-7m) (k-m) (k+m)^2 {1\over \mathcal{J}^4} \cr & +{1\over 256}
( k - m ) \left( 15k^5 - 135k^4m + 182k^3m^2 - 94k^2m^3
                               + 171km^4 - 11m^5 \right) {1\over \mathcal{J}^6}
+O\left({1\over \mathcal{J}^8}\right) \,.
\ea
\ee
We shall now combine these terms and obtain the energy shifts up to
third order in perturbation theory.


\subsection{First and Second order}

The first and second order terms in the $1/\mathcal{J}^2$ expansion of the energy shift (\ref{full}) are
\begin{equation}
\begin{aligned}
\delta E_1^{osc}
&= \sum_n {2(k-m)m - n^2 + n\sqrt{n^2 + 4 m(m-k)} \over 2}
\cr
\delta E_2^{osc}
&= \sum_n - {n (-4m \left( k - m \right) \left( 5k^2
  - 15km + 6m^2 \right)  + 2\left( k - 3m \right) \left( 3k - 2m
  \right) n^2 + n^4)  \over 8 \sqrt{n^2 + 4 m(m-k)}} \cr
& \qquad + {1\over 16}(
\left( k - m \right)  {\left( k + m \right) }^2 \left( -3 k + 7 m
\right)  + 4 \left( 3 k^2 - 10 k m + 5 m^2 \right)  n^2 + 2 n^4) \,.
\end{aligned}
\end{equation}
The large $n$ behaviour of the summand in $\delta E_1^{osc}$ is
$1/n^2$, which ensures that the energy shift at first order is
finite. In the second order term the summand has asymptotics
\begin{equation}\label{Anomaly}
(\delta E_{2}^{osc})_n= -{1\over 16} (k-m)^2 (3k^2 - 14 km + 19 m^2)+
  O\left({1\over n^2}\right)\,.
\end{equation}
Thus, there is an anomalous pieces, which needs to be regularized.
Applying zeta-function regularization the regularized energy reads
\begin{equation}
\begin{aligned}
&\left(\delta E_2^{osc}\right)_{reg}
  =  {1\over 32}(k-m)^2 (3k^2 - 14 km + 19 m^2)  \cr
& +\sum_n \left\{{n (-4m \left( k - m \right) \left( 5k^2 - 15km +
  6m^2 \right)  + 2\left( k - 3m \right) \left( 3k - 2m \right) n^2 +
  n^4)
 \over  8 \sqrt{n^2 + 4 m(m-k)}} \right.\cr
&\left.\qquad \quad  + {1\over 8}(-2 \left( k - m \right)  m \left( 4
  k^2 - 11 k m + 3 m^2 \right)  + 2 \left( 3 k^2 - 10 k m + 5 m^2 \right)  n^2 + n^4)
  \right\}
\,.
\end{aligned}
\end{equation}
Combining the zero-mode energy shift with the oscillator contribution,
we obtain in summary that at order $1/\mathcal{J}^2$ and
$1/\mathcal{J}^4$ the shift is
\begin{equation}\label{TwoLoopStringFinal}
\begin{aligned}
&\delta E_1^{string} =  {1\over 2} m (k-m)+ \sum_n {2(k-m)m - n^2 + n\sqrt{n^2
      + 4 m(m-k)} \over 2} \cr
&\delta E_2^{string} =  -{1\over 8}m (k-m) (4k^2 - 11 km + 3 m^2)  \cr
& +\sum_n \left\{- {n (-4m \left( k - m \right) \left( 5k^2 - 15km + 6m^2 \right)  + 2\left( k - 3m \right) \left( 3k - 2m \right) n^2 + n^4)
 \over  8 \sqrt{n^2 + 4 m(m-k)}} \right.\cr
&\left.\qquad \quad
+ {1\over 8}(
-2 \left( k - m \right)  m \left( 4 k^2 - 11 k m + 3 m^2 \right)  + 2 \left( 3 k^2 - 10 k m + 5 m^2 \right)  n^2 + n^4)
\right\} \,.
\end{aligned}
\end{equation}


\subsection{Third order}

Further expanding the string theory result for the contributions of
the oscillators to the energy up to third order, \ie, order $1/\mathcal{J}^6$, yields
\be\label{ThreeLoopOsc}
\ba
\delta E_3^{osc}
       =& \sum_n {1\over 128} \{15k^6 - 150k^5m + 317k^4m^2 -
        276k^3m^3 + 265k^2m^4 - 182km^5 + 11m^6 \cr
            &+
            4( 15k^4 + 4k^3m - 36k^2m^2 - 64km^3 + 65m^4 ) n^2 - 8( 15k^2 - 38km +
        19m^2 ) n^4 - 8n^6\} \cr
        & + {1\over 16 (n^2+ 4m (m-k))^{3/2}}\big\{
            4( k - m )^2 m^2( 45k^4 - 324k^3m +
            621k^2m^2 - 370km^3 + 60m^4 ) n \cr
        & - 2( k - m ) m( 53k^4 - 481k^3m +
            1083k^2m^2 - 815km^3 + 192m^4 ) n^3 \cr
        &  +
            ( 15k^4 - 218k^3m + 603k^2m^2 - 556km^3 +
            164m^4 ) n^5 + ( 15k^2 - 44km + 25m^2 )n^7 + n^9 \big\} \,.
\ea
\ee
The sum is again divergent as
the large $n$ behaviour of the summand in (\ref{ThreeLoopOsc}) is
\be\label{EthreeAsymp}
\ba
\left(\delta E_3^{osc}\right)_n
       &= {9\over 32} ( k - m )^2( 5k^2 -
       18km + 17m^2 ) n^2 \cr
        &+  {1\over 128} ( k - m )^2( 15k^4 -
    248k^3m + 766k^2m^2 - 752km^3 + 91m^4) + O\left({1\over n^2}\right)\,.
\ea
\ee
We again apply zeta-function regularization. In the present case,
we need to evaluate the Riemann zeta function $\zeta (s)=
\sum_{n=1}^\infty 1/n^s$ at $s=-2, 0$. The values can be calculated by
writing the zeta-function as
\be
\zeta (s) = {1\over 1-2^{1-s}}
            \sum_{n=0}^\infty {1\over 2^{n+1}} \sum_{k=0}^n (-1)^k
        {n\choose k} (k+1)^{-s} \,,
\ee
and evaluating the inner sum first. This results for $k>1$ in
\be
\zeta (-k+1) = -{B_k \over k}\,,
\ee
where $B_k$ are the Bernoulli numbers. Now $B_3=0$ and
therefore only $\zeta (0)$ gives a non-vanishing contribution, namely
$\zeta (0) = -1/2$.
The regularized contribution from the oscillators to the zero modes is thus
\be
(\delta E_3^{osc})_{reg} = {1\over 256}
( k - m)^2\left(15k^4 - 248k^3m + 766k^2m^2 - 752km^3 + 91m^4 \right)
+ \sum_n \cdots \,,
\ee
where the dots indicate the non-zero mode contributions, with the
terms in (\ref{EthreeAsymp}) subtracted.

Combining all terms, we arrive at the third order energy shift as
computed from the string theory side
\be\label{StringThreeLoop}
\ba
\delta E_3^{string} &={1\over 16} ( k - m ) m( 8k^4 - 52k^3m + 89k^2m^2 - 42km^3 + 5m^4)\cr
        & +\sum_n {1\over 16} \{
                   2 (k-m)m ( 8k^4 - 52k^3m + 89k^2m^2 - 42km^3 + 5m^4) \cr
        & - ( 15k^4 -128 k^3m + 279k^2m^2 - 202km^3 + 44m^4 ) n^2 - ( 15k^2 - 38km +
             19m^2 ) n^4 - n^6\} \cr
        & + {1\over 16 (n^2+ 4m (m-k))^{3/2}}\big\{
            4( k - m )^2 m^2( 45k^4 - 324k^3m +
            621k^2m^2 - 370km^3 + 60m^4 ) n \cr
        & - 2( k - m ) m( 53k^4 - 481k^3m +
            1083k^2m^2 - 815km^3 + 192m^4 ) n^3 \cr
        & +
            ( 15k^4 - 218k^3m + 603k^2m^2 - 556km^3 +
            164m^4 ) n^5 + ( 15k^2 - 44km + 25m^2 )n^7 + n^9 \big\} \,.
\ea
\ee
We shall see subsequently, that this regularized energy shift agrees
with the prediction from the Bethe ansatz.

\setcounter{subsection}{0}

\appendix{Details of Bethe ansatz computation}\label{BetheDetails}




\subsection{Zero-modes}

The zero mode integral is
\be
\delta E^{(0)} = {c\over k }\oint_{{\cal C}_{ab}} dz {f'(z)\over (z-c)} \,.
\ee
By deforming the contour to infinity, we pick up the residues at $z=c$ and
$z=\pm 1$.

Combining these residues and subsequently expanding them
in $1/\mathcal{J}$ by making use of (\ref{cESol}), yields
\be\label{BetheFinalZero}
\ba
\delta E^{(0)}
&={1\over 2} m(k-m){1\over \mathcal{J}^2}\cr
&-{1\over 8} m(k - m)( 4k^2 - 11km + 3m^2 )
                   {1\over \mathcal{J}^4}\cr
&+{1\over 16}( k - m )m( 8k^4 - 52k^3m + 89k^2m^2 - 42km^3 + 5m^4)
                               {1\over \mathcal{J}^6}
+ O\left({1\over \mathcal{J}^8}\right)
\,.
\ea
\ee
Comparison to the string theory result, which were computed
in the previous section shows that up to third order in the
$1/\mathcal{J}^2$ perturbation expansion, the zero-mode terms
(\ref{BetheFinalZero})
agree with the ones of the zeta-function regularized expressions on the string side.


\subsection{Non-zero modes}
\label{nonZM}

The non-zero mode contributions come from the sum in (\ref{CotExp}) and are
\be\label{EOscRes}
\ba
\delta E^{osc}
= \sum_{n=1}^\infty \delta E^{(n)}
=& {2 c\over k}  \sum_{n=1}^\infty
\oint_{{\cal C}_{ab}} dz {f'(z) f^2(z) \over (z-c)  (f^2(z) -n^2)}  \,.
\ea
\ee
Again, deforming the contour to infinity, we pick up (possibly
non-trivial) residues at $z=c$, $z=\infty$, $z=\pm 1$ as well as
$z=z_n$, where $z_n$ were defined in (\ref{fzn}).

The residues at $z=c$ and $z=\infty$ vanish.
The residue at $z=z_n$ was evaluated in (\ref{Reszn}).
In order to expand this in $1/\mathcal{J}$, one first needs to solve
(\ref{pofz}) perturbatively for $z_n$ (note that there are two
roots $z_n$ each for positive $n$ and for negative $n$).

The expansion of (\ref{Reszn}) yields up to third order
\be\label{ZNRes}
\ba
&\hbox{Res}_{z=z_n}  \cr
&= {1\over 2}
    \bigg\{2 k(m-k)- n^2 + n\sqrt{n^2+ 4m(m-k)} \bigg\}  {1\over \mathcal{J}^2} \cr
&+ {1\over 8}
    \bigg\{-2k(m-k)(k^2-11km + 6m^2)  + 2( 3k^2 - 10km + 5m^2 ) n^2 + n^4 \cr
&   \qquad -{-4( k - m )m( 5k^2 - 15km +6m^2)n + 2( k - 3m )(3k - 2m)n^3 + n^5
            \over  \sqrt{n^2 + 4 m (m-k)}}
    \bigg\}  {1\over \mathcal{J}^4} \cr
&+ {1\over 32}
    \bigg\{-9k^6+184k^5m-848k^4m^2 + 1380k^3m^3 -934k^2m^4 + 252km^5 -20m^6 \cr
&   \qquad - 2(15k^4-128k^3m+279k^2m^2-202km^3+44m^4)n^2-2(15k^2-38km +19m^2)n^4 - 2n^6\cr
&   \qquad +  {2 \over (n^2 + 4 m (m-k))^{3/2}}
    \bigg(
          4(k-m)^2m^2( 45k^4 - 324k^3m + 621k^2m^2 - 370km^3 + 60m^4) n   \cr
&   \qquad -2( k - m) m( 53k^4 - 481k^3m + 1083k^2m^2 - 815km^3 + 192m^4) n^3  \cr
&   \qquad +( 15k^4 - 218k^3m + 603k^2m^2 - 556km^3 + 164m^4) n^5 \cr
&   \qquad       +(
      15k^2 - 44km + 25m^2) n^7 + n^9 \bigg)
\bigg\}  {1\over \mathcal{J}^6}\cr
&+ O\left({1\over \mathcal{J}^8}\right)\,.
\ea
\ee

Finally, there are the residues at $z=\pm 1$, which contribute to
the $n$-independent terms of the summands $\delta E^{(n)}$
\be\label{PMOneRes}
\ba
&\hbox{Res}_{z=1} + \hbox{Res}_{z=-1}\cr
& \qquad = (k^2-m^2){1\over \mathcal{J}^2} - {1\over 4} (k - m )(k+m)(k^2- 8km+3m^2)
               {1\over \mathcal{J}^4}\cr
& \qquad -{1\over 32}k ( -9k^5 + 152k^4m - 608k^3m^2 +
               816k^2m^3 - 410km^4 + 64m^5 )
                               {1\over \mathcal{J}^6}
+ O\left({1\over \mathcal{J}^8}\right)
\,.
\ea
\ee
Putting the residues in (\ref{ZNRes}) and (\ref{PMOneRes}) together we
obtain
\be\label{BetheFinalOsc}
\ba
&\delta E^{(n)} \cr
&= {1\over 2}
    \bigg\{-2m(m-k)- n^2 + n\sqrt{n^2+ 4m(m-k)} \bigg\}  {1\over \mathcal{J}^2} \cr
&+ {1\over 8}
    \bigg\{ -2(k-m) m(4k^2 - 11km +3m^2)  + 2(3k^2 - 10km + 5m^2)n^2 + n^4 \cr
&   \qquad -{-4( k - m )m( 5k^2 - 15km +6m^2)n + 2( k - 3m )(3k - 2m)n^3 + n^5
            \over  \sqrt{n^2 + 4 m (m-k)}}
    \bigg\}  {1\over \mathcal{J}^4} \cr
&+ {1\over 32}
    \bigg\{ 4(k-m)m(8k^4 - 52k^3m +89k^2m^2 - 42km^3+5m^4)   \cr
&   \qquad -2(15k^4 - 128k^3m + 279k^2m^2-202km^3+44m^4)n^2
           -2(15k^2 - 38km + 19m^2) n^4 - 2n^6 \cr
&   \qquad + {2 \over (n^2 + 4 m (m-k))^{3/2}}
    \bigg(
          4(k-m)^2m^2( 45k^4 - 324k^3m + 621k^2m^2 - 370km^3 + 60m^4) n   \cr
&   \qquad -2( k - m) m( 53k^4 - 481k^3m + 1083k^2m^2 - 815km^3 + 192m^4) n^3  \cr
&   \qquad +( 15k^4 - 218k^3m + 603k^2m^2 - 556km^3 + 164m^4) n^5 \cr
&   \qquad +(
      15k^2 - 44km + 25m^2) n^7 + n^9 \bigg)
\bigg\}  {1\over \mathcal{J}^6} \cr
& + O\left({1\over \mathcal{J}^8}\right)\,.
\ea
\ee
The complete energy shift is then
\be\label{BetheFinalEnergy}
\delta E = \delta E^{(0)} + \sum_{n=1}^\infty \delta E^{(n)}\,,
\ee
where the various terms are written out in (\ref{BetheFinalZero}) and
(\ref{BetheFinalOsc}).

In summary, the Bethe result agrees with the string results
(\ref{TwoLoopStringFinal}), (\ref{StringThreeLoop}) including
order $1/\mathcal{J}^6$.



\appendix{Details of the large $k$ string computation}\label{LargekDetails}

We evaluate the energy shift $\delta E^{string}$ in the large $k$ limit, for fixed
$m$ and ${\mathcal J}$. Again, the problematic part in the computation
are the $\omega$-dependent terms, for which
we are forced to use approximations for finding the roots
in different regions of the parameters.

Note that first expanding the summands in
(\ref{ZeroModeE}) and (\ref{StringEnergyShift}) $1/k$ before summing
them yields divergent expressions. However, unlike the divergences
that occured in the $1/\mathcal{J}$ expansion at second and third
order in perturbation theory, these divergences cannot be removed,
using standard regularisation procedures such as zeta-function
regularisation as they contain logarithmic divergences. The origin of
this divergence is the irregular dependence on $k$ of the resummed
expression (\ref{dEosc}).

In order to ascertain what kind of function we are summing, it is
useful to numerically plot the summands. This is done in figure 3 for various,
mainly large, values of $k$.
\begin{figure}
\begin{center}
\includegraphics*[width=.22\textwidth]{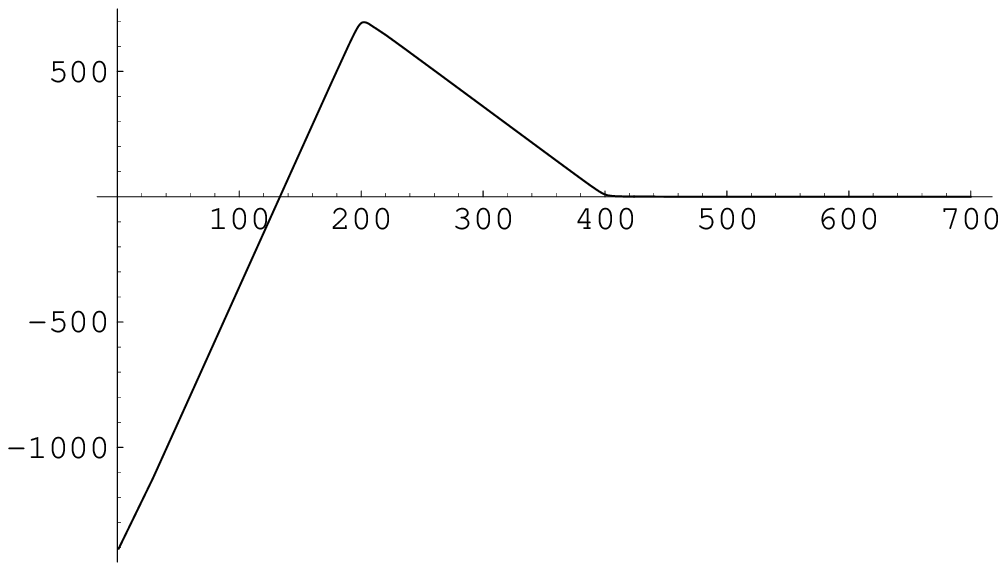}
\quad
\includegraphics*[width=.22\textwidth]{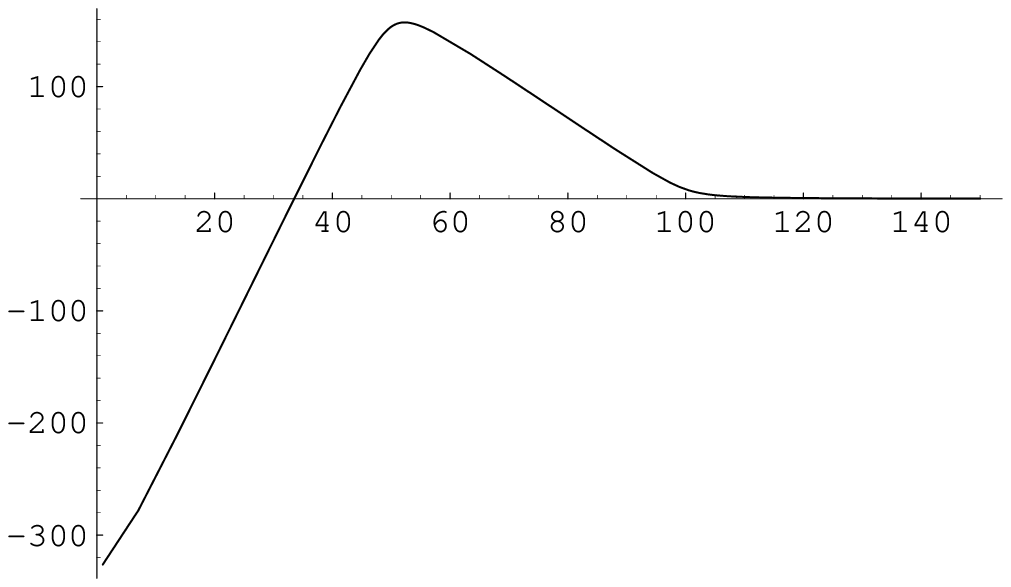}
\quad
\includegraphics*[width=.22\textwidth]{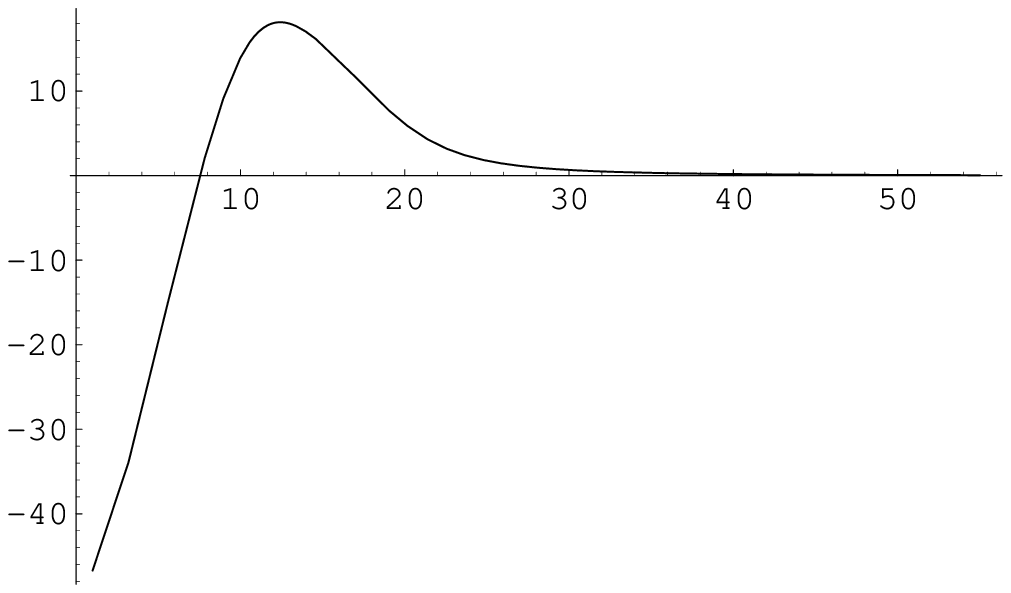}
\quad
\includegraphics*[width=.22\textwidth]{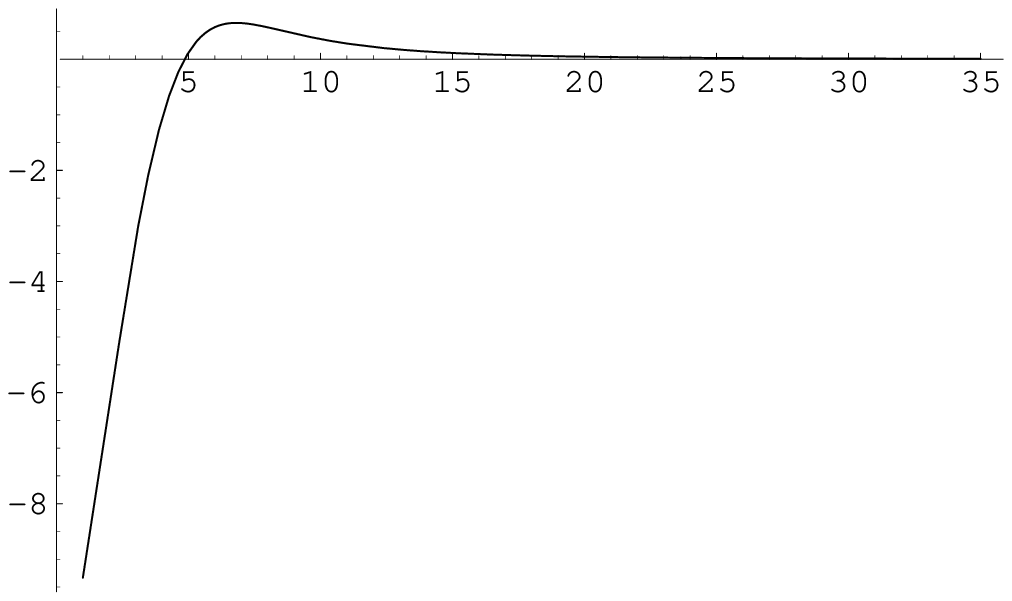}
\caption{Profiles of the summands for $k=400$, $k=100$, $k=20$ and
  $k=5$, respectively, with $({\mathcal J}= 3$, $m=2)$.}
\end{center}
\end{figure}
\begin{figure}[t]
\begin{center}
\includegraphics*[width=.22\textwidth]{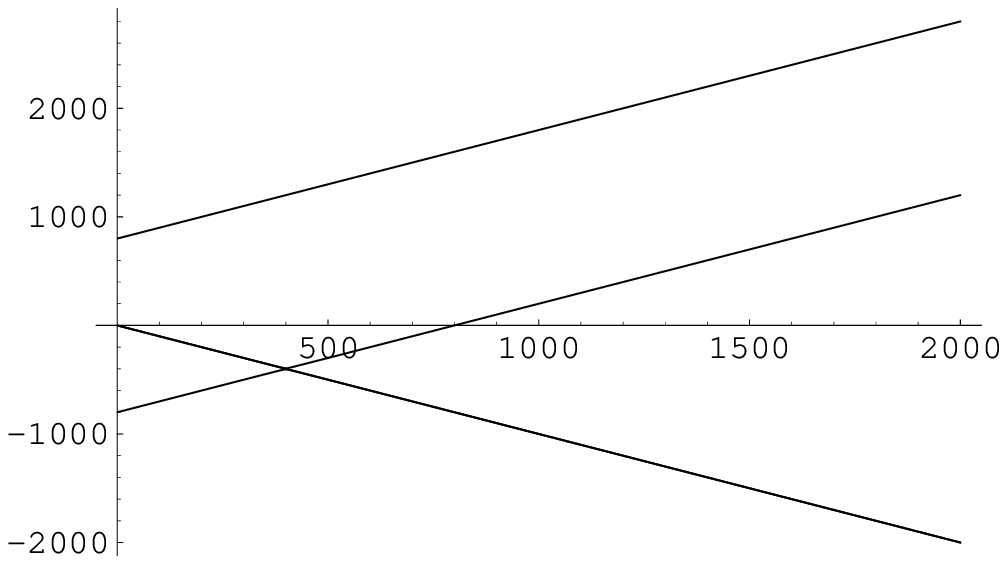}
\quad
\includegraphics*[width=.22\textwidth]{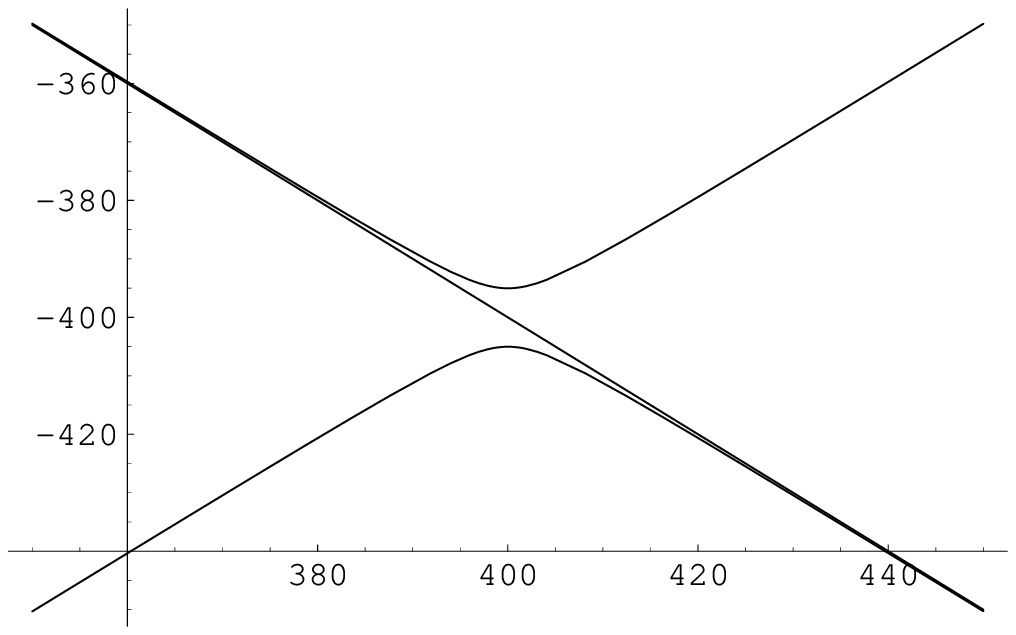}
\end{center}
\caption{Profiles of the $\omega$ frequencies  for $k=400,
 ({\mathcal J}= 3$, $m=2$). The plot on the right hand side zooms into
 the plot on the left hand side.}
\label{f:omega}
\end{figure}
Solving (\ref{quartic1}) in the limit $n\sim |k|\rightarrow \infty $
we find, up to $O(1/k^2)$ corrections
\begin{eqnarray}\label{o12}
 \omega _{n\,1,2}&=&n\pm 2|k|\pm \frac{1}{|k|}\left[
  m\mathcal{J}+\frac{n\pm 2|k|}{n\pm |k|}\,\,
  \frac{(\mathcal{J}+m)^2}{2}
 \right],
 \\ \label{o34}
 \omega _{n\,3,4}&=&-n-\frac{(\mathcal{J}+m)n}{2|k|(n^2-k^2)}
 \left[
 (\mathcal{J}+m)|k|\pm\sqrt{(\mathcal{J}-m)^2n^2+4m\mathcal{J}k^2}
 \right].
\end{eqnarray}
These expressions approximate the frequencies well enough in the
entire range of $n$, except for $n-|k|=O(1)$, where $1/k$ corrections
to $\omega _2$ and $\omega _3$ blow up. Solving (\ref{quartic1}) in
that region we find
\begin{eqnarray}\label{}
 \omega _1&=&3|k|,
 \\
 \omega _4&=&-n,
 \\ \label{opm}
 \omega _{\pm}&=&-|k|\pm\sqrt{(n-|k|)^2+(\mathcal{J}+m)^2}\,.
\end{eqnarray}
Comparing (\ref{opm}) to (\ref{o12}), (\ref{o34}) we see that
$\omega _+$ asymptotes $\omega _2$ at $n\gg |k|$ and $\omega _3$ at
$n\ll |k|$, while $\omega _-$ asymptotes $\omega _3$ at $n\gg |k|$
and $\omega _2$ at $n\ll |k|$. Thus $\omega _2$ and $\omega _3$
interchange at $n=|k|$ by passing through the singularity.

Computing the sign factors from (\ref{signs}) we get
\begin{equation}
\begin{aligned}
\label{}
&n<|k|:\qquad  &sign C_1^{(n)}&=&1,&&
 sign C_2^{(n)}&=&-1, &&
 sign C_3^{(n)}&=&1, &&
 sign C_4^{(n)}&=&-1,&&
 \\
&n>|k|:\qquad &sign C_1^{(n)}&=&1, &&
 sign C_2^{(n)}&=&1, &&
 sign C_3^{(n)}&=&-1, &&
  sign C_4^{(n)}&=&-1,&&
  \\
&n-|k|\sim 1:\qquad  &sign C_1^{(n)}&=&1, &&
 sign C_4^{(n)}&=&-1, &&
 sign C_\pm^{(n)}&=&\pm 1\,.
\end{aligned}
\end{equation}
We are now ready to compute the sum over modes. To do that we divide
the range of summation into three parts
\begin{eqnarray}
&{\rm (I)}&1\leq n\leq |k|-s-1  \nonumber \\
&{\rm (II)}&|k|-s\leq n\leq |k|+s  \nonumber   \\
&{\rm (III)}& |k|+s+1\leq n\,,
\end{eqnarray}
where $1\ll s\ll |k|$. In the regions (I) and (III) the summation of $O(1/k)$
terms can be replaced by an integration over $x=n/|k|$
\begin{eqnarray}\label{}
 &&\sum_{n=1}^{\infty }\sum_{I=1}^{4}\left(sign C_I^{(n)}\omega _{n,I}
 -n-\frac{\kappa ^2}{2n}\right)=
 4\sum_{n=1}^{|k|-s-1}(|k|-n)-2(\mathcal{J}+m)^2\sum_{n=1}^{|k|-s-1}
 \frac{1}{n}
 \nonumber \\
 &&+\int_{0}^{1-s/|k|}dx\left[
 2m\mathcal{J}+(\mathcal{J}+m)\frac{(2-x^2)(\mathcal{J}+m)+x
 \sqrt{(\mathcal{J}-m)^2x^2+4m\mathcal{J}}}{1-x^2}
 \right]
 \nonumber \\
 &&+\sum_{l=-s}^{s}\left[
 2\sqrt{l^2+(\mathcal{J}+m)^2}-3l
 \right]
 +2(\mathcal{J}+m)^2\int_{1-s/|k|}^{\infty }dx\left(
 \frac{x}{x^2-1}-\frac{1}{x}
 \right)
 \nonumber \\
 &&=2\left[k^2-|k|+(\mathcal{J}+m)\sqrt{m\mathcal{J}}+m\mathcal{J}
 +F(0,\mathcal{J}+m)+(\mathcal{J}+m)^2\ln\frac{\sqrt{\mathcal{J}+m}}
 {\sqrt{\mathcal{J}}+\sqrt{m}}\right].
\end{eqnarray}
Combining this with the expansion of the zero modes
\be
\delta E^{(0)} = - {|k| \over {\mathcal J} + m } + \bigg( 1 + 2
  \sqrt{ {{\mathcal J} - m \over {\mathcal J} + m} } \bigg) -
 \,
 { 7  {\mathcal J}^2 + 10 {\mathcal J} m + m^2 \over 2 \, |k|
  ({\mathcal J} + m)}\,  +
  {\mathcal O} ({1\over k^2}) \,,
\ee
and (\ref{dEosc}) we obtain
\begin{eqnarray}\label{}
 \delta E&=&\frac{k^2-4\gamma ^2+m\mathcal{J}+
 2F\left(0,\sqrt{\mathcal{J}^2-m^2}\right)
 +2F\left(0,\mathcal{J}+m\right)
 -4F\left(\{\frac{|k|}{2}\},\sqrt{\mathcal{J}
 (\mathcal{J}+m)}\right)}{\mathcal{J}+m}
 \nonumber \\
 &&+\sqrt{m\mathcal{J}}+(\mathcal{J}+m)\ln\frac{\sqrt{\mathcal{J}+m}}
 {\sqrt{\mathcal{J}}+\sqrt{m}}\,.
\end{eqnarray}
Since $\gamma =|k|/2+O(1/k)$, this expression has a finite $k\rightarrow \infty
$ limit, as was observed numerically. In order to determine the
asymptotic values
of the constant one needs  the expression for $\gamma $ with an $O(1/k)$ accuracy
\begin{equation}
\gamma = {|k| \over 2} + { m (2 {\mathcal J} + m) \over 4 |k|} +
         O\left({1\over k^3} \right) \, ,
\end{equation}
which implies
\begin{eqnarray}\label{}
 \delta E&=&\frac{
 2F\left(0,\sqrt{\mathcal{J}^2-m^2}\right)
 +2F\left(0,\mathcal{J}+m\right)
 -4F\left(\{\frac{|k|}{2}\},\sqrt{\mathcal{J}
 (\mathcal{J}+m)}\right)}{\mathcal{J}+m}
 \nonumber \\
 &&+\sqrt{m\mathcal{J}}+(\mathcal{J}+m)\ln\frac{\sqrt{\mathcal{J}+m}}
 {\sqrt{\mathcal{J}}+\sqrt{m}}\, - m \, .
\end{eqnarray}
For large enough $\alpha$, the function $F(\alpha, \beta)$ can be approximated as in (\ref{approximation}), and thus the previous sum can be further simplified to
\begin{eqnarray}\label{}
 \delta E&=&
- {1\over 2} ({\mathcal J} + m) \ln({\mathcal J} + m) - ({\mathcal J} - m)\ln ({\mathcal J} -m ) + 2 {\mathcal J}\ln {\mathcal J} +
\nonumber \\
 &&+\sqrt{m\mathcal{J}}- (\mathcal{J}+m)\ln
 (\sqrt{\mathcal{J}}+\sqrt{m})\, - m \, +  O  \left( {1\over \alpha} \right) \, .
\end{eqnarray}

\newpage

\end{document}